\documentclass[a4paper,12pt]{article}
\usepackage{graphicx}
\usepackage{epsfig}
\usepackage{jcappub}
\renewcommand{\arraystretch}{2}

\begin{document}

\title{Post-Planck Dark Energy Constraints}
\author[a]{Dhiraj Kumar Hazra}
\author[b]{Subhabrata Majumdar}
\author[c]{Supratik Pal}
\author[d]{Sudhakar Panda}
\author[e]{Anjan A. Sen}

\affiliation[a]{Asia Pacific Center for Theoretical Physics, Pohang, Gyeongbuk
790-784, Korea}
\affiliation[b]{Tata Institute for Fundamental Research, Mumbai,400005, India}
\affiliation[c]{Physics and Applied Mathematics Unit, Indian Statistical
Institute, Kolkata, 700108, India}
\affiliation[d]{Harish Chandra Research Institute, Allahabad,211019, India}
\affiliation[e]{Center For Theoretical Physics, Jamia Millia Islamia, New
Delhi-110025, India}

\emailAdd{dhiraj@apctp.org, subha@tifr.res.in, supratik@isical.ac.in}
\emailAdd{panda@mri.ernet.in, aasen@jmi.ac.in}

\abstract{
We constrain plausible dark energy models, parametrized by multiple candidate
equation of state, using the
recently published Cosmic Microwave Background (CMB) temperature anisotropy data from Planck together with the WMAP-9
low-$\ell$ polarization data and data from low redshift surveys.
To circumvent the limitations of any particular equation of state towards 
describing all existing dark energy models, we work with 
three different equation of state covering a broader class of dark energy models and, hence, provide more robust and
generic constraints on the dark energy properties. 
We show that a clear tension exists between dark energy constraints from CMB and non-CMB observations when one allows for
dark energy models having both phantom and non-phantom behavior; while CMB is
more
favorable to phantom models, the low-z data prefers model with behavior close to a Cosmological Constant. Further, we reconstruct the
equation of state of dark energy as a function of redshift using the results
from combined CMB and non-CMB data and find that Cosmological Constant
lies outside the 1$\sigma$ band for multiple dark energy models allowing phantom behavior. A considerable fine tuning is needed to keep
models with strict non-phantom history inside 2$\sigma$ allowed range. This
result might motivate one to construct phantom models of dark energy,
which is achievable in the presence of higher derivative operators as in string theory.
However, disallowing phantom behavior, based {\it only} on strong theoretical
prior, leads to both CMB and non-CMB datasets agree on the nature of
dark energy, with the mean equation of state being very close to the
Cosmological Constant. Finally, to illustrate the impact of additional dark energy parameters on other 
cosmological parameters, we provide the cosmological parameter constraints for the 
``Standard Model of Cosmology'' including an evolving dark energy component, for the different dark energy models. }

\maketitle

\section{Introduction}
It is now established beyond doubt by a range of cosmological observations,
that
our universe is going through a late time accelerated expansion phase.
To explain such an accelerating universe, one either needs to add an additional
exotic component, called dark energy, in the energy budget of the universe that
necessarily has a negative pressure causing an overall repulsive behavior of
gravity at large cosmological scales (see \cite{review} for some excellent
reviews),
or one has to modify Einstein's General Relativity. Unfortunately,
while candidates for
dark energy have been proposed, its exact nature remains unknown. Alternatively, satisfactory
modifications of General Relativity consistent with gravitational physics at
astrophysical scales are also lacking.

Tight observational evidence is still lacking as to whether dark energy, where
one has to add an extra component in the energy budget,  has a constant energy
density throughout the history of the universe (termed as Cosmological
Constant), or if it evolves in time. Specifically, if it evolves, one would like to know
its equation of state which governs this evolution. One would also like to know if this
equation of state satisfies the weak energy condition such that it dilutes with the cosmological expansion. 
Moreover whether dark energy violates the weak energy condition and behaves like some mysterious
form of phantom energy with its energy density increasing with time and possibly leading to  a future singularity
is also discussed widely in literatures. 

The majority of the current and future cosmological observational programs are
dedicated
to finding answers to these pertinent questions. These include, among others, 
(i) the construction of the Hubble diagram using Supernova Type-Ia as
standard candle \cite{sn},  (ii) measuring the tiny
fluctuations present in the temperature of the cosmic microwave background
radiation \cite{Planck,komatsu}, or (iii) measuring the oscillations present
in the matter power spectrum through large scale structure surveys
\cite{Eisenstein}.

The simplest example for dark energy is the Cosmological Constant ($\Lambda$).
The
concordance $\Lambda$CDM model is consistent with most of the
cosmological observations. However, deep theoretical issues such as fine tuning
as well
cosmic coincidence problems have motivated people to explore
beyond the Cosmological Constant, the natural alternatives being
scalar field models. A variety of such scalar field models, including
string theory embeddings for a positive Cosmological Constant~\cite{Kachru:2003sx},
quintessence
\cite{quint}, k-essence \cite{kess}, phantom fields \cite{phant}, tachyons
\cite{tach} etc have been proposed. 
Other than scalar field models, a barotropic fluid with an equation of
state $p(\rho)$, such as the Generalized
Chaplygin Gas (GCG) and its various generalizations \cite{gcg,ss1} have
also been considered for dark energy model building.

Since current cosmological observations provide us a precise description of
the universe and impose tight constraints on the standard
cosmological model, the general
behavior of the dark energy component is also constrained. On
the other hand,
given the proliferation of dark energy models in the literature, it is not
practical to confront each model
with the observational data. Rather one needs to look for generic 
features of dark energy that are present in
a large class of models and then to confront these features with the
observational results. The most popular
way of doing this is to assume a parametrization for the dark energy equation of state $w$ as a function redshift, $z$,
or the scale factor, $a$. However, such parametrization should describe a wide
variety of
dark energy models so that by constraining
a minimal set of parameters, one can constrain a large set of representative
dark energy models.
One such widely used parametrization
is the Chevallier-Polarski-Linder (CPL) parametrization first discussed by
Chevallier and Polarski~\cite{Chevallier:2000qy} and later by
Linder~\cite{Linder:2002et}.
It uses a linear dependence of the equation of state on the scale factor and
contains two parameters. This parametrization
has been used by  all recent cosmological observations, including Planck, to put
constraints on dark energy. However,  since the
CPL parametrization has the dark energy equation of state as a linear
dependence on $a$, it may not represent models with more complicated $a$
dependence at slightly higher redshifts where dark energy contribution might
still be
non-negligible. Hence, constraining dark energy behavior using {\it only} the
CPL
parametrization might give biased, or even incorrect, conclusions.

Given the fact that Planck~\cite{Planck}, in combination with
non-CMB observations like SN-Ia, BAO, HST etc, has measured the cosmological
content of
the universe with unprecedented accuracy, it is now interesting
to investigate how different parametrizations of dark energy equation-of-state
can result in different constraints
on dark energy behavior when confronted with the observational
data. We can, now, ask the following questions - Are conclusions regarding the
nature of dark energy borne out of observational data biased by our choice of
parametrization? Or does a general pattern exist in the dark energy behavior
that is always
true even if we consider different parametrizations for dark energy equation of
state?

In this paper, we investigate these issues by considering two more
parametrizations for the dark energy equation of state together with the CPL
parametrization. The first of the two parametrizations used in this paper was
proposed by Scherrer and Sen (SS)~\cite{Scherrer:2007pu} and it represents
slow-roll thawing class of canonical scalar
field models having an equation of state which varies very close to the $w=-1$
irrespective of the form of the potential. Subsequently, the idea was also
extended to phantom type scalar field models with a negative kinetic energy term\cite{ss2} where it was shown the
parametrization holds true also for such scalar field models. Recently, it was
also
shown~\cite{Ali:2009mr} that this parametrization can
represents scalar field models with DBI type kinetic energy term under
similar slow-roll conditions.

The second parametrization which we consider, was proposed by Bento, Bertolami
and
Sen~\cite{gcg} and subsequently was
discussed~\cite{Bento:2002uh,Bento:2002yx,Bento:2003dj,Bento:2003we}
for more general parameter ranges by Scherrer and Sen \cite{ss1} and is known as Generalized Chaplygin Gas (GCG) parametrization. In this
parametrization, for a certain choice of parameter range, the dark energy
equation of state behaves like a
thawing class of scalar field models where the present acceleration is a
transient one. For a different choice of parameter range, the parametrization
represents the freezing/tracker type models.
Hence with a single equation of state parametrization,
one can model both the thawing as well as freezing class of scalar field models. 
Recently, this parametrization has been used~\cite{Thakur:2012rp}
to study the Bayesian Evidence for thawing/freezing class of the dark energy
models using different observational results including WMAP-7 results. 

For all of the three parametrizations, we reconstruct the redshift evolution of
the dark energy equation of state. For each case, we also investigate the
departure of the cosmological parameters from the $\Lambda$CDM best-fit values.Note, that the CPL parametrization has
already been discussed in the Planck analysis~\cite{Planck:2013kta}. However our analysis, apart from
showing a consistency check with Planck results, provides some new facts and
highlights a tension between CMB and non-CMB
observations. Using
the CMB , non-CMB and combined data of both, we show a clear tension between
high-redshift and low redshift measurements for models allowing phantom
behavior.
Moreover from the combined 
analysis, using the correlation between the equation of state parameters we
reconstruct the allowed range of dark energy evolution with redshift and address the stand of phantom and non-phantom models in the allowed band. In particular,
we show that once we allow phantom behavior, the allowed non-phantom
behavior is extremely close to the Cosmological Constant, $w=-1$, behavior.

At this point it should be noted that, as an alternative approach, model
independent reconstructions of
the expansion history ({\it i.e.} the Hubble parameter $h(z)$) have been
carried out before~\cite{Shafieloo:2007cs,Shafieloo:2012yh}, from which the
equation of state can also be reconstructed. Also note
that
certain parametrization of the dark energy equation of state may 'artificially'
limit the
properties of dark energy from explaining a few effects,
such as the recent slowdown of cosmic acceleration~\cite{Shafieloo:2009ti}.
To overcome such biases, in this current work, we use three different kinds of
dark energy parametrizations in order to
cover a broad spectrum of dark energy behavior. We constrain the nature of dark
energy using CMB
and non-CMB surveys, taken individually as well as jointly (see \cite{recent}
for some recent works on dark energy constraint after Planck). 

%

The paper is organized as follows: in 
Section 2, we briefly describe the three dark energy 
parametrizations that we use in this paper; in Section 3, we describe the
different observational
datasets that we use to constrain the dark energy evolution;
Section 4 describes the results and we conclude in Section 5.

\section{Dark Energy Parametrizations}
\vspace{2mm}
\subsection{CPL Parametrization}

This parametrization, first proposed by Chevallier and Polarski and later
reintroduced by Linder has the following form:

\begin{equation}
w(a)= w_{0} + w_{\rm a} (1-a) = w_{0} + w_{\rm a}\frac{z}{1+z},~\label{eq:cpleq}
\end{equation}

\noindent
where $w_{0}$ and $w_{\rm a}$ are the two parameters in the model. They represent
the equation of state at present ($a=1$) and its
variation with respect to scale factor (or redshift). From the
infinite past till the present time, the equation of state is bounded between
$w_{0}+w_{\rm a}$ and $w_{0}$. The dark energy density, in this case, evolves as:

\begin{equation}
\rho_{\rm DE} \propto a^{-3(1+w_{0}+w_{\rm a})} e^{-3 w_{\rm a} (1-a)}. 
\end{equation}

\noindent
This equation of state remarkably fits a wide range of scalar field dark energy models including  the supergravity-inspired SUGRA dark
energy models.
The CPL parametrization is most commonly used in the literature 
to study the nature of dark energy. An outcome of the specific form of this
parametrization is that, for
$w_{0} \geq-1$ and $w_{\rm a} > 0$, the dark energy remains non-phantom in nature
throughout
the cosmological evolution; otherwise it shows phantom behavior at some point in time.

\vspace{2mm}

\subsection{SS Parametrization}

This parametrization was proposed by Scherrer and Sen~\cite{Scherrer:2007pu} for slow-roll `thawing' class of scalar field models having a canonical kinetic
energy
term.
Later, it was shown~\cite{Ali:2009mr} that this parametrization also holds
true for the tachyon-type scalar field models having DBI-type kinetic
energy term \cite{pandadbi} as well as for phantom models for scalar fields
having negative kinetic energy term. The main motivation for this
parametrization was to look for a unique dark energy evolution for scalar field
models that are constrained to evolve very close
to the Cosmological Constant ($w=-1$). Since similar situations also arise in 
inflationary scenarios in the early universe, one assumes the same slow-roll
conditions on the scalar field potentials
as used in inflation. However, unlike the inflationary epoch, the situation
differs due to the presence of
the large matter content in the late universe; still it can be shown that under
the
assumption of the (i) two slow-roll conditions and that (ii) 
the scalar field is initially frozen at $w=-1$ due to large Hubble damping
(termed
`thawing class'), one gets a unique form for the dark energy equation of state
irrespective of its potential. The form of this equation of state, for a
universe with flat spatial hypersurface, is given by:

\begin{eqnarray}
w(a) &=& (1+w_{0})\left[\sqrt{1+(\Omega_{\rm DE}^{-1}-1)a^{-3}}-(\Omega_{\rm
DE}^{-1}-1)a^{-3} \tanh^{-1}\frac{1}{\sqrt{1+(\Omega_{\rm
DE}^{-1}-1)a^{-3}}}\right]^2\times \nonumber\\
&\times& \left[\frac{1}{\sqrt{\Omega_{\rm DE}}} - \left(\frac{1}{\Omega_{\rm
DE}} - 1\right) \tanh^{-1} \sqrt{\Omega_{\rm DE}}\right]^{-2} -1.
\end{eqnarray}

\noindent
This parametrization has one model parameter, $w_{0}$, which represents its
value
at the present epoch together with
the general cosmological parameter $\Omega_{\rm DE}$ representing the present
day dark energy density, which
in a flat universe is related to the present day matter energy density through
$\Omega_{\rm m} + \Omega_{\rm DE} = 1$.
The energy density of this model of dark energy can be calculated
analytically using the Friedmann equations which has been used in our analysis.
At this point, it is worth pointing out that the recently constructed axionic
quintessence
model in string theory \cite{pst} can be described by this parametrization for
certain range of parameters.

\subsection{GCG Parametrization}

The Chaplygin gas (CG) equation of state was first discussed in the cosmological context by Kamenschik {\it et. al.}~\cite{Kamenshchik:2001cp}
and is described by 

\begin{equation}
p = - \frac{c}{\rho},
\end{equation} 

\noindent
where $c$ is an arbitrary constant and $p$ and $\rho$ represents the pressure
and energy density of the CG fluid. Subsequently,
this equation of state was generalized by Bento et al \cite{gcg} and Billic et
al \cite{bilic} as

\begin{equation}
p = - \frac{c}{\rho^{\alpha}},
\end{equation} 

\noindent
where $\alpha$ is a constant within the range $0\leq\alpha\leq1$. This form is
known as the Generalized Chaplygin Gas (GCG) equation of state. In a
later work Scherrer and Sen \cite{ss1} considered the parameter range $\alpha <
0$ to describe diverse cosmological behaviors. Assuming a
spatially homogeneous and isotropic universe along with the energy momentum
conservation
equation gives us

\begin{equation}
w(a) = -\frac{A}{A + (1-A) a^{-3(1+\alpha)}},~\label{eq:gcgeq}
\end{equation}

\noindent
where $A = c/\rho_{\rm GCG}^{1+\alpha}$. The GCG parametrization also contains
two model parameters e.g $A$ and $\alpha$,
similar to the CPL parametrization.

It is easy to check that the present equation of state is $w(0)\,=\,-A$. For
$(1+\alpha) > 0$, $w(a)$ behaves like
a dust in the past and evolves towards negative values and becomes $w=-1$ in the asymptotic future. This is similar to `tracker/freezer'
behavior for a scalar field where it tracks the background matter in the past,
and in the late time behaves like a dark energy with
a negative equation of state. For $(1+\alpha) < 0$, the opposite happens. In
this
case the $w(a)$ is frozen to $w=-1$ in the past and the slowly
evolves towards higher values and eventually behaves like a dust in the future.
This behavior is similar to the `thawing' class of scalar field
models. Moreover, in this case, the late time acceleration is a
transient phenomena as the acceleration slows down eventually and the Universe
enters again a dust regime.
For both 'thawing' or 'freezer' kind of behaviors, the transition to/from $w=-1$
depends on the value of $\alpha$.
We shall consider both $1+\alpha >0$ and $1+\alpha <0$ to
 consider freezing as well as thawing type behaviors. 
However, we restrict $0<A<1$ only since for $A>1$ singularity appears at
finite past. As a consequence, this model is restricted to probe dark energy
behavior for non-phantom cases only, a feature which is also
true for scalar field models with positive kinetic energy. This is akin
to a strong theoretical prior on the model.

To summarize, we are considering three parametrization for the dark energy
equation of state. The CPL parametrization has a simple linear dependence on
scale factor which is
true for dark energy behavior in general around present day but may not
represent models that have more complicated scale factor dependence at slightly
higher redshifts. The SS
parametrization represents all slow-roll thawing class of scalar field models
even if they are phantom in nature (i.e., having negative kinetic energies).The
parametrization is
also applicable for scalar fields with canonical kinetic energies as well as
non-canonical
kinetic energies. Hence, if dark energy is not Cosmological
Constant but has small deviation from the Cosmological Constant, the SS is a
good
parametrization to consider. Moreover, the deviation from Cosmological Constant
can be represented by a single parameter. Our third parametrization is the GCG
parametrization which can describe both thawing as well as the freezer type
behaviors of the scalar field models. It, however, restricts us to consider only
non-phantom models. This parametrization can represent dark energy behaviors
where the acceleration of the Universe
have slowed down recently.

Finally, let us emphasize that while the CPL parametrization was proposed as a
phenomenological form for the equation of state of dark energy, both the SS and
GCG parametrizations were obtained from a specific field theory Lagrangian under certain conditions \cite{gcg,ss2}.

\section{Methodology}

In this paper we have put constraints on the late time evolution
history of dark energy by contrasting multiple dark energy models with CMB and
low redshift observations. For CMB
we have used the recent Planck $C_{\ell}^{\rm TT}$ data. As Planck has not yet
released the observed Polarization data, we have used
WMAP-9~\cite{Hinshaw:2012aka} low-$\ell$ polarization data for completeness, as
has also been used in Planck analysis. In different frequency channels
Planck has detected the CMB sky in much smaller scales ($\ell=2500$) compared to WMAP. Planck has published two likelihood estimators~\cite{Planck:2013kta},
namely, the low-$\ell$ (2-49) likelihood which is estimated by {\tt commander}
and the high-$\ell$ (50-2500) which is estimated by {\tt CAMspec}, for four different
frequency channels. At small scales, the foreground effects are dominant and
there are 14 nuisance parameters~\cite{Planck:2013kta,Ade:2013zuv} corresponding to the foreground effects in different frequency channels. For our analysis 
with CMB data, we have always marginalized over these nuisance parameters. The
cosmological model is described by the six cosmological parameters:
$\Omega_{\rm b}h^2$, $\Omega_{\rm CDM}h^2$, $\theta$, $\tau$, $A_{\rm S}$ and
$n_{\rm S}$. First four parameters describe the background where $\Omega_{\rm b}$
and $\Omega_{\rm CDM}$ represent the baryon and the cold dark matter density and $h$ represents the Hubble parameter. $\theta$ is the ratio of the sound horizon
to the angular diameter distance at decoupling and $\tau$ is the reionization
optical depth. $A_{\rm S}$ and $n_{\rm S}$ describes the amplitude and the
spectral index of the primordial perturbation which we assume to be of the power-law form. Finally, the CPL and GCG model have 2 parameters and SS model have one
parameter
describing the dark energy equation of state.

For non-CMB data, we have used Supernovae data, Baryon Acoustic Oscillations
(BAO) data
and data from Hubble space telescope. For Supernovae data
we have used the recent Union 2.1 compilation~\cite{Suzuki:2011hu} with 580
supernovae within redshifts $\sim0.015-1.4$. We have used the covariance
matrix of Union 2.1 compilation which includes systematic errors. 
For the Baryon Acoustic Oscillations we have used four datasets, namely the 
six-degree field galaxy survey~\cite{Beutler:2011hx}, SDSS
DR7~\cite{Percival:2009xn} and
BOSS DR9 measurements~\cite{Anderson:2012sa} and the data from WiggleZ
survey~\cite{Blake:2011en}.  We confront the theoretical model
with the
distance ratio ($d_z=r_{s}(z_{\rm drag})/D_{\rm V}(z)$) measured by 
the particular surveys (and with the functions of $d_z$), where $z_{\rm drag}$ is the particular redshift where
the
baryon-drag optical depth becomes unity and $r_{s}(z_{\rm drag})$ is the
comoving
sound horizon at that redshift. $D_{\rm V}(z)$ is related to the angular
diameter distance and the Hubble parameter at redshift $z$. For BAO we get
constraints from 6 data points, of which three come from WiggleZ survey at 3
different
redshifts ($z=0.44,\,0.6,\,0.73$) and the other three come from the two SDSS
measurements such as SDSS DR7 : $z=0.35$, SDSS DR9 : $z=0.57$ and a 6DF :
$z=0.106$).
We have also used the HST data~\cite{Riess:2011yx} which uses the nearby type Ia Supernova data with Cepheid calibrations to
constrain the value of $H_0$.

We have used the publicly available cosmological Boltzmann code
{\tt CAMB}~\cite{Lewis:1999bs,cambsite} to calculate the power spectrum for CMB and
different observables for non-CMB
observations and for the Markov Chain Monte Carlo analysis we have used the
{\tt CosmoMC}~\cite{Lewis:2002ah,cosmomcsite}.  In all the cases, we have taken into account 
the effect of dark energy perturbations though {\tt CAMB}. Throughout our analysis
we have fixed the number of relativistic species to be $N_{\rm eff}=3.046$.

\section{Results}

\begin{table*}[t]
  
\begin{center}
  \hspace*{0.0cm}\begin{tabular}
  {|c|c|c|c|c|} \hline 
  Data & $\Lambda$CDM & CPL & GCG & SS \\ \hline    

{Planck (low-$\ell$ + high-$\ell$)} & 7789.0 & 7787.4 & 7789.0 & 7788.1 \\
\hline

{WMAP-9 low-$\ell$ polarization} & 2014.4 & 2014.436 & 2014.383& 2014.455\\
\hline

  {BAO : SDSS DR7}	&   0.410 &0.073  &0.451 & 0.265\\  \hline

  {BAO : SDSS DR9}	& 0.826 &  0.793 & 0.777 & 0.677  \\  \hline

  {BAO : 6DF}	& 0.058 & 0.382 &  0.052 & 0.210 \\  \hline
  
  {BAO : WiggleZ}	&   0.020&  0.069 & 0.019 &   0.033 \\  \hline

  {SN  : Union 2.1}		&545.127 &546.1  & 545.131 &545.675 \\  \hline

  {HST}		& 5.090 & 2.088 &  5.189 & 2.997 \\  \hline\hline 

  {Total}		&10355.0 & 10351.4 & 10355.0 &10352.4 \\  \hline \hline 

  \end{tabular}
  \end{center}
\caption{~\label{tab:chi2}Best fit $\chi_{\rm eff}^2$ obtained for different
models upon comparing against CMB + non-CMB datasets. The breakdown of
the $\chi_{\rm eff}^2$ for individual data are also provided. To obtain the
best-fit we have used the Powell's BOBYQA method of iterative minimization.}
\end{table*}

We start this section by showing, in Table~\ref{tab:chi2}, the best-fit
$\chi^{2}_{\rm eff}$ values for cosmological models with
different dark energy parametrizations. In obtaining the
 best-fits, we have used Powell's BOBYQA method of iterative
minimization~\cite{powell} and the $\chi^{2}_{\rm eff}$ quoted are
obtained from the joint analysis with CMB and all non-CMB data described in the
previous section. We
have also shown the breakdown of the $\chi^{2}_{\rm eff}$
from various datasets to clarify the preference of any individual dataset
towards different
dark energy models. Notice that the 
GCG parametrization, which is a non-phantom model, has
similar $\chi^2_{\rm eff}$ values as in $\Lambda$CDM although it contains two
extra parameters. This shows that if we restrict ourselves to
non-phantom models - {\it a strong theoretical prior} -  it is hard to 
distinguish them from a $\Lambda$CDM behavior as the best-fit value is
always close to a model with $w=-1$. Allowing phantom behavior (i.e., for both
CPL and SS
parametrizations) results in marginally better fit
to the complete datasets by a $\Delta \chi^{2}_{\rm eff} \sim{\cal O}(2-4)$. The improvement from $\Lambda$CDM for CPL model is 3.6 and for SS model is 2.6. This improvement in $\chi^{2}_{\rm eff}$ is, arguably, not large
enough to justify the necessity to venture into the phantom regime. 

It is clear from Table~\ref{tab:chi2},  where the total and the individual $\chi^2_{\rm eff}$ are given, that the Supernova data marginally favors the concordance $\Lambda$CDM model
as well as the non-phantom GCG model 
compared to other two parametrizations. 
 Planck data prefers lower $H_0$ for $\Lambda$CDM which does not agree well with the $H_0$ from HST and the fit to HST gets worse. 
However CPL and SS models allowing phantom equation of state fit the data from both CMB and HST better than cosmological constant at their best fit values. 
A direct upshot of these phantom equation of state that fits the CMB data better than the $\Lambda$CDM model is that it comes with higher value of $H_0$ that in turn agrees with the HST data better too.  
However, for the same choice of best fit parameter values they do not provide similar improvement in $\chi^{2}_{\rm eff}$ for Supernovae data. This indicates that 
SN data favors the equation of states closer to the cosmological constant.
The breakdown of $\chi^{2}_{\rm eff}$  points to the tantalizing possibility that
different datasets, which effectively probe different epochs in the history of our Universe,  
prefer different kind of dark energy
behaviors.  The improvement in the total $\chi^2_{\rm eff}$ for the CPL and SS model over $\Lambda$CDM and GCG is driven by the ability to have a higher $H_0$ albeit along with the associated preference for phantom behavior. However, to be conclusive, one would need to evaluate the full likelihood behavior for the different models w.r.t the diverse datasets.

\begin{figure*}[!t]
 
\begin{center} 
\resizebox{180pt}{140pt}{\includegraphics{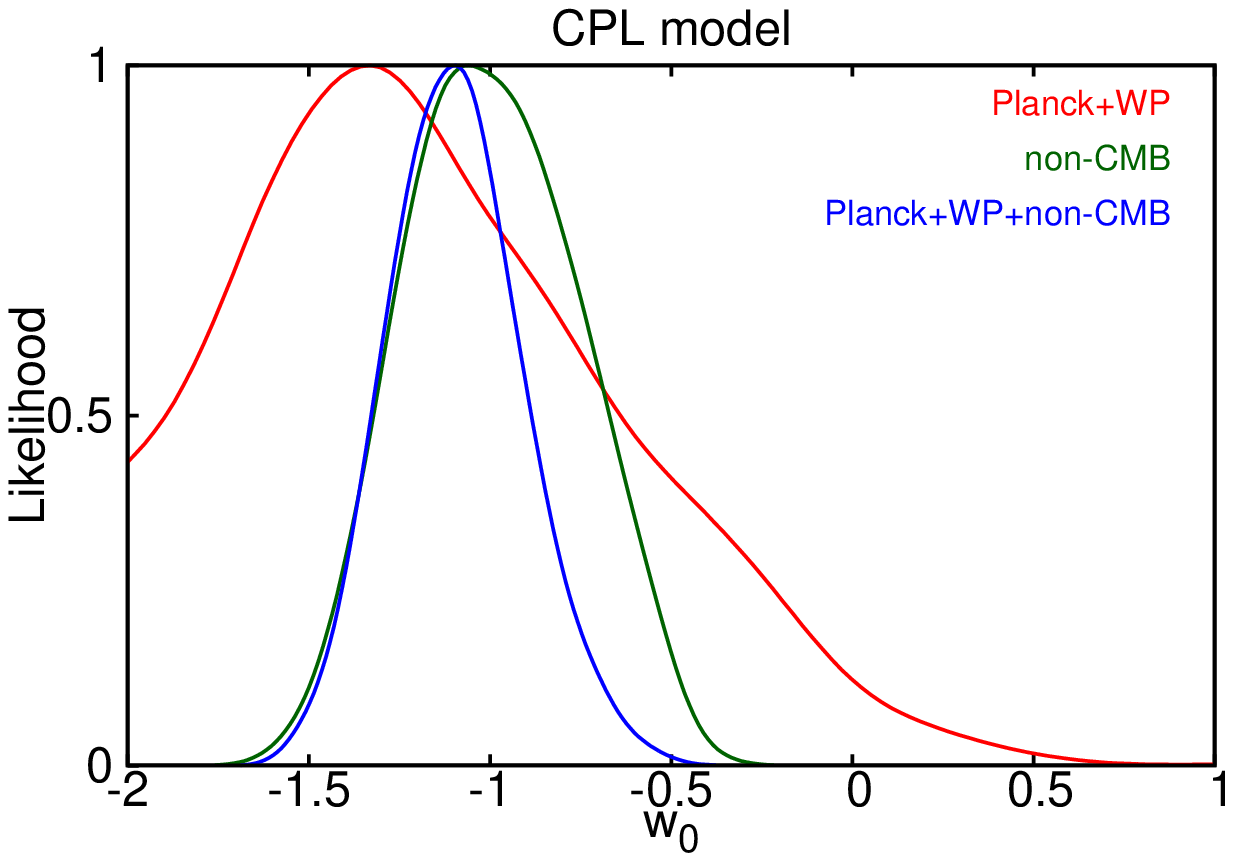}} 
\resizebox{180pt}{140pt}{\includegraphics{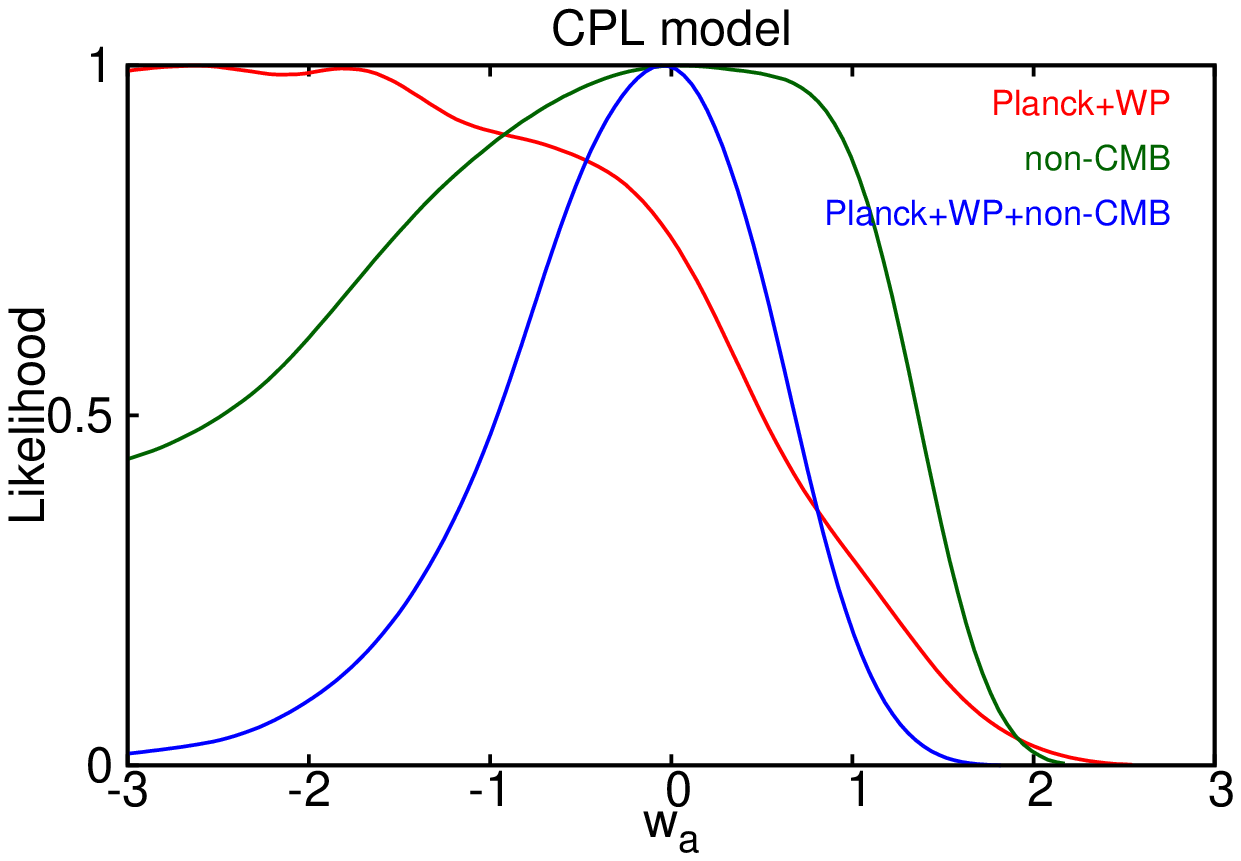}} 
\resizebox{180pt}{140pt}{\includegraphics{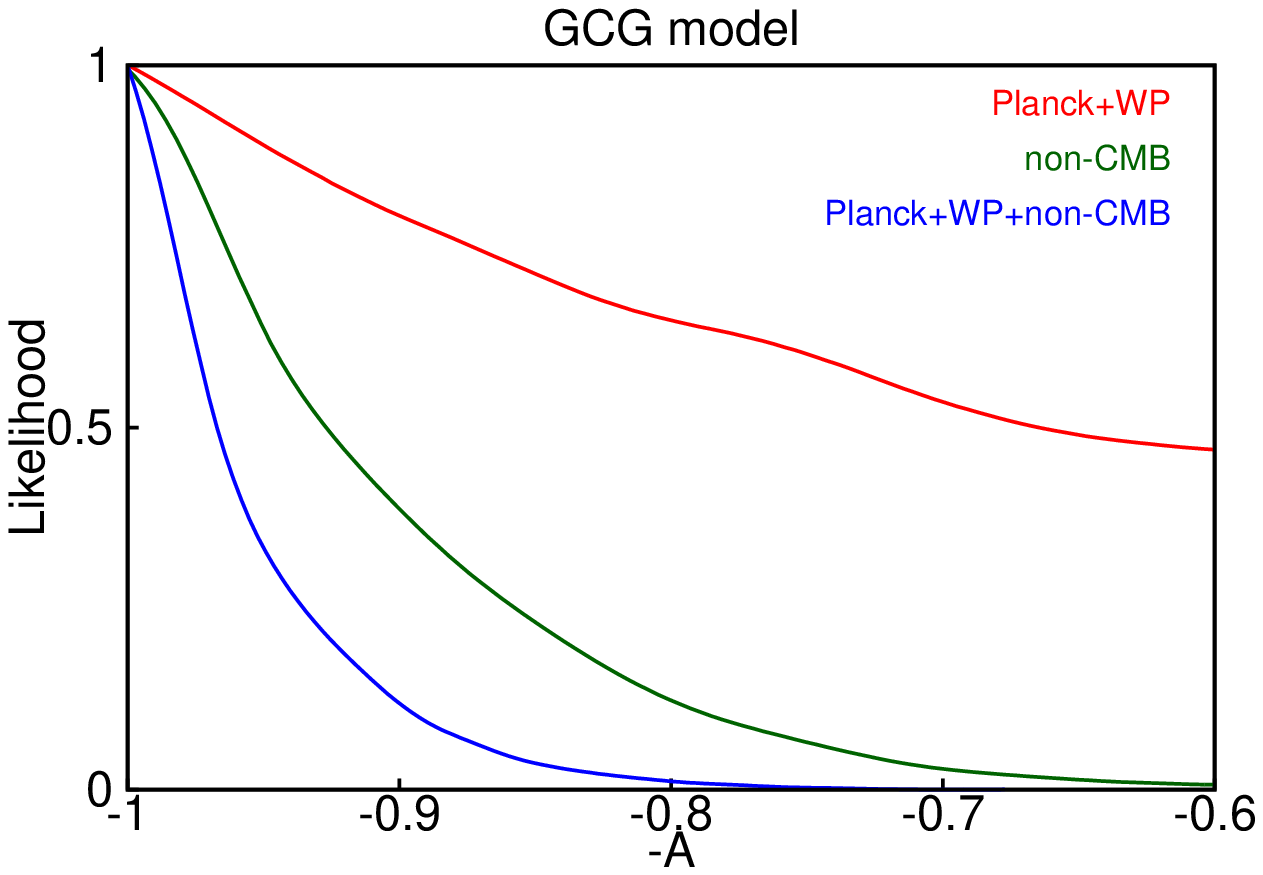}} 
\resizebox{180pt}{140pt}{\includegraphics{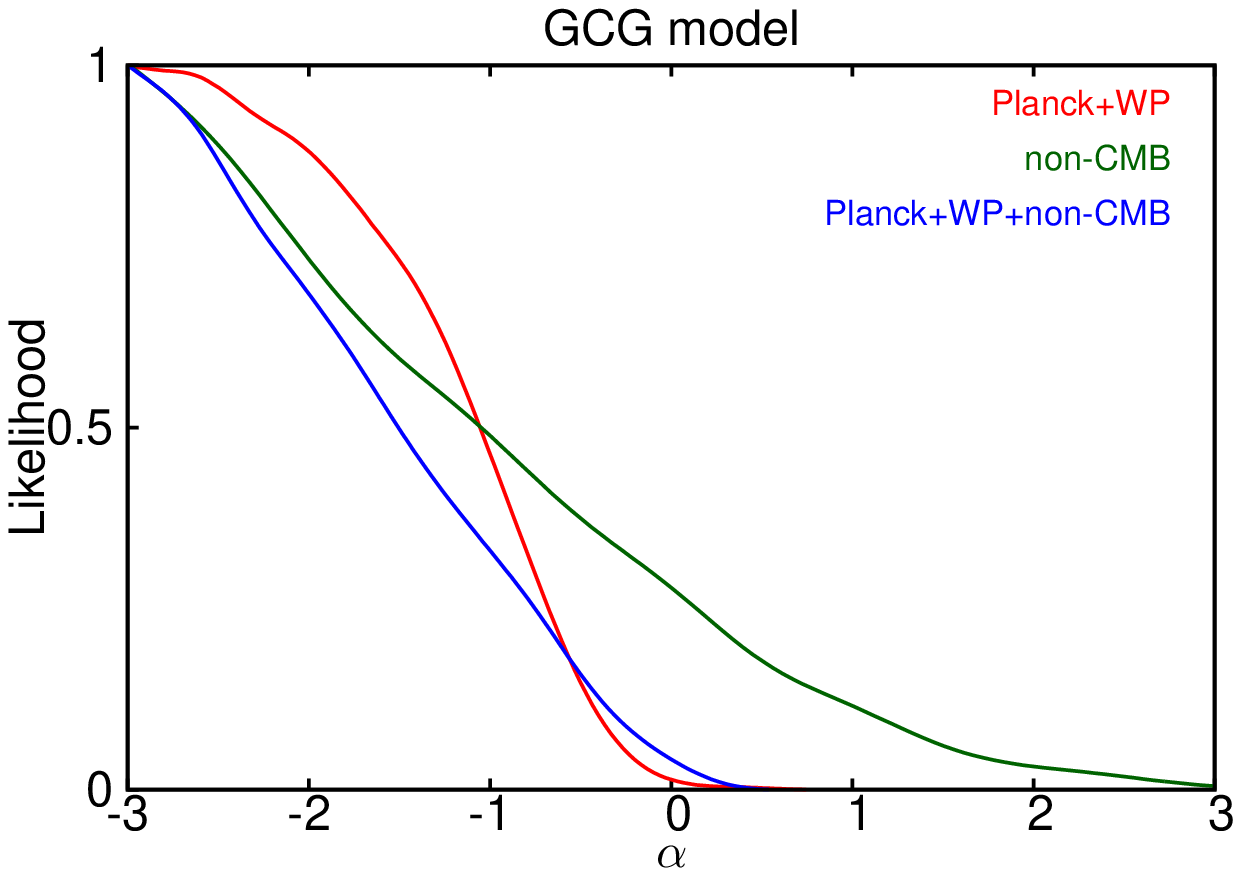}} 
\resizebox{180pt}{140pt}{\includegraphics{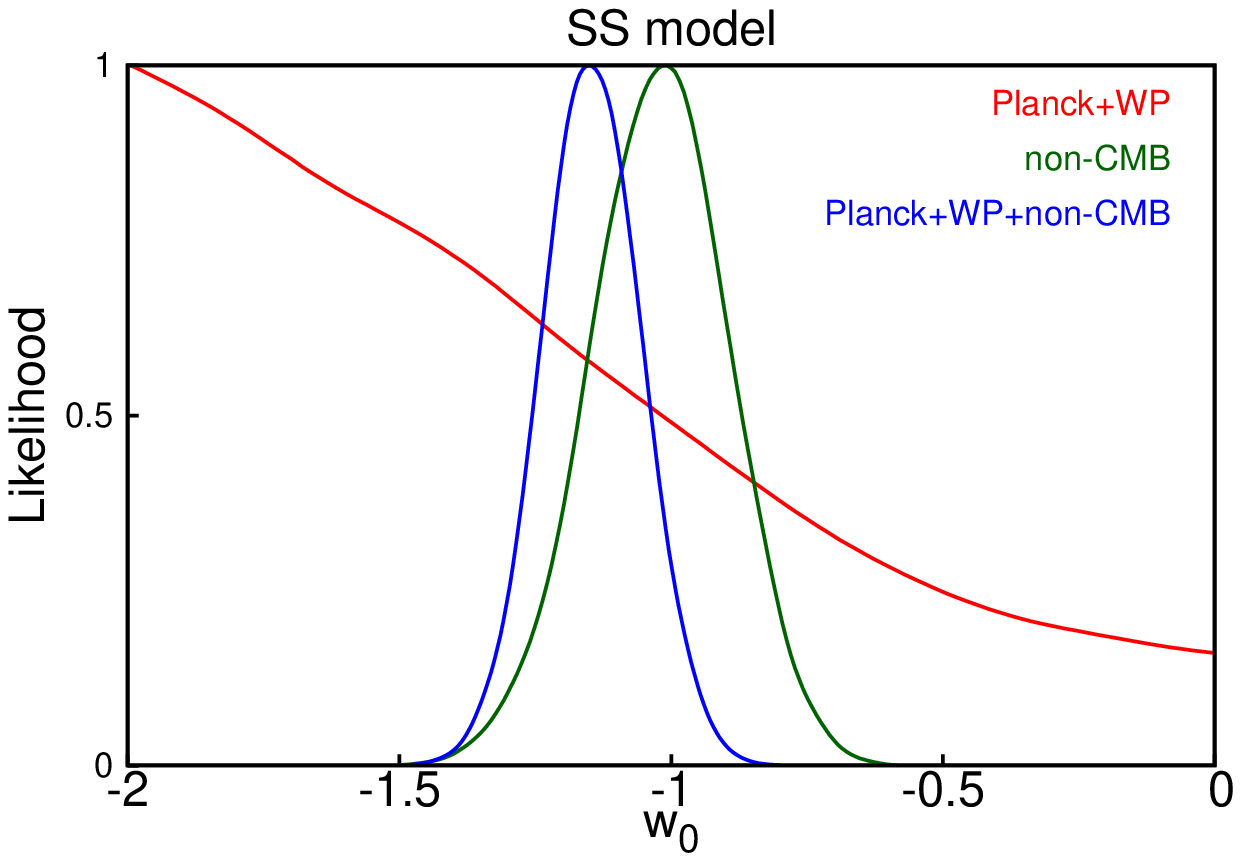}} 

\end{center}
\caption{~\label{fig:1dcntrs}The likelihood functions for different parameters
of equation of state. The upper ones are for the CPL parametrization, the middle ones
for the GCG parametrization and the bottom one is for the SS parametrization. The
color codes are for different analysis with different observational data and
are described in the plot.}
 \end{figure*}


\renewcommand{\arraystretch}{1.1}

\begin{table*}[!htb]
\begin{center}
\vspace{4pt}
\begin{tabular}{|c | c | c | c |}
\hline\hline
 & CPL& SS &GCG\\

\hline

 & $0.0221\pm0.00028$ & $0.0221\pm0.00026$ &
$0.022\pm0.00028$ \\

$\Omega_{\rm b}h^2$ & $0.022\pm0.00026$ & $0.0221^{+0.00026}_{-0.00024}$ &
$0.0223\pm0.00024$ \\

 & $0.027^{+0.004}_{-0.005}$ & $0.028^{+0.004}_{-0.006}$ &
$0.029\pm0.005$ \\

\hline
 & $0.1196\pm0.0027$ & $0.1198\pm0.0026$ &
$0.1199^{+0.0026}_{-0.0028}$ \\

$\Omega_{\rm CDM}h^2$ & $0.1209\pm0.0023$ & $0.1192\pm0.0018$ &
$0.117\pm0.0015$ \\

 & $0.126^{+0.014}_{-0.017}$ & $0.128^{+0.014}_{-0.018}$ &
$0.127^{+0.015}_{-0.018}$ \\

\hline

 & $1.041\pm0.0006$ & $1.041\pm0.0006$ &
$1.041\pm0.0006$ \\

$100\theta$ & $1.041\pm0.0006$ & $1.041\pm0.00056$ &
$1.042\pm0.00056$ \\

 & $1.042\pm0.023$ & $1.048\pm0.022$ &
$1.05^{+0.019}_{-0.027}$ \\

\hline
 & $0.09^{+0.012}_{-0.014}$ & $0.09^{+0.012}_{-0.015}$ &
$0.09^{+0.013}_{-0.014}$ \\

$\tau$ & $0.087^{+0.012}_{-0.014}$ & $0.091\pm0.013$ &
$0.094\pm0.014$ \\

 & - & - &- \\

 \hline
 & $-1.13^{+0.37}_{-0.66}$ & $-1.31^{+0.19}_{\rm unbounded}$ &
$-0.827^{+0.06}_{\rm Non-phantom~prior~cut}$ \\

$w_0 [-A]$ & $-1.005^{+0.15}_{-0.17}$ & $-1.14^{+0.08}_{-0.09}$ &
$-0.957^{+0.007}_{\rm Non-phantom~prior~cut}$ \\

 & $-0.995^{+0.23}_{-0.27}$ & $-1.02\pm0.12$ &
$-0.92^{+0.018}_{\rm Non-phantom~prior~cut}$ \\

\hline
 & $-1.15^{+0.6}_{\rm unbounded}$ 
 & - & $-1.97^{+0.32}_{\rm unbounded}$\\

$w_{\rm a}$[$\alpha$]& $-0.48^{+0.77}_{-0.54}$ & - & $-2.0^{+0.29}_{\rm unbounded}$\\

 & $-0.5^{+1.64}_{-0.94}$ & - & $-1.49^{+0.4}_{\rm unbounded}$\\

\hline
 & $0.9607\pm0.007$& $0.9603\pm0.007$ &
$0.9603\pm+0.00073$ \\

$n_{\rm S}$ & $0.9579^{+0.0063}_{-0.0066}$ & $0.9619^{+0.0059}_{-0.0057}$ &
$0.9669^{+0.00056}_{-0.00059}$ \\

& -& -&- \\

\hline
& $3.089^{+0.023}_{-0.027}$ & $3.089^{+0.023}_{-0.028}$ &
$3.09\pm0.025$ \\

$\ln[10^{10}A_{\rm S}]$ & $3.087^{+0.024}_{-0.026}$ & $3.091\pm0.025$ &
$3.092\pm0.026$ \\

& -& -&- \\

\hline\hline
& $0.239^{+0.028}_{-0.099}$& $0.27^{+0.04}_{-0.1}$&
$0.344^{+0.022}_{-0.032}$\\

$\Omega_{\rm m}$ & $0.291^{+0.011}_{-0.013}$& $0.288^{+0.012}_{-0.013}$&
$0.304^{+0.009}_{-0.011}$\\

& $0.29\pm0.024$& $0.298^{+0.02}_{-0.026}$&
$0.3^{+0.021}_{-0.024}$\\
\hline

& $80^{+17.8}_{-7.8}$& $74.8^{+13.3}_{-9.8}$& $64.6^{+2.61}_{-1.91}$\\

$H_{0}$ & $70.26\pm1.4$& $70.3\pm1.4$& $67.9^{+0.9}_{-0.7}$\\

& $72.68\pm2.2$& $72.67\pm2.15$& $72.4\pm2.16$\\
\hline\hline
\end{tabular}
\end{center}
\caption{~\label{tab:param}The mean values and the $1\sigma$ ranges for different cosmological parameters 
from our analysis. CPL, SS and GCG marks the dark energy parametrization used. 
The parameters $w_0$ and $w_a$ represents $-A$ and $\alpha$ for GCG model and have been indicated in the
table. For each parameters the first, second and last row indicate the results from the analysis with 
CMB, CMB + non-CMB and non-CMB datasets respectively.}

\end{table*}

In Table~\ref{tab:param}, we quote the mean values as well as the $1\sigma$ errors bars for different parameters 
($\Omega_{\rm b}h^2$, $\Omega_{\rm CDM}h^2$, $\theta$, $\tau$, $n_{\rm S}$, $A_{\rm S}$, $w_0$, $w_{\rm a}$ and the derived matter 
content $\Omega_{\rm m}$ and Hubble constant $H_0$). For each parameter, the first row contains results from 
CMB datasets, the second row contains the results when we combine high and low redshift data and the last row provides 
the results from non-CMB datasets. Note that bounds on $H_0$ becomes considerably weak for CMB analysis alone when we allow
phantom equation of state through CPL and SS. In all the cases $w_0$ ($-A$ for GCG) is better constrained with non-CMB data compared to the CMB 
data alone since $w_0$ determines the behavior of dark energy model at low redshift. However, $w_{\rm a}$ ($\alpha$ for GCG) 
is constrained better by CMB datasets for CPL and GCG as it determines the change in the dark energy equation of state throughout the expansion of Universe. 
It is clear that CPL and SS allow lower $\Omega_{\rm m}$ and higher $H_{0}$ whereas the GCG allows higher 
$\Omega_{\rm m}$ and lower $H_{0}$ consistently for all the analyses~\footnote{For analysis with non-CMB datasets GCG indicates a higher 
value of $H_0$}. It should be noted that for CPL model the mean value of $H_{0}$ for the analysis with CMB datasets 
comes out to be $80$ and the 1$\sigma$ uncertainty stretches the upper bound to 98. For SS model the bound on 
$H_{0}$ CMB {\it only} analysis is better than CPL since the model has one dark energy parameter and therefore less degenerate.

Having described the best-fit $\chi^2_{\rm eff}$ and the allowed ranges for
different parameters, we now discuss the likelihoods of parameters for the different dark energy parametrization. These likelihoods are shown in  Figure~\ref{fig:1dcntrs}.
 Interestingly, in CPL as well SS parametrization, the CMB data from Planck takes the
present value of equation of state ($w_{0}$) towards higher
phantom values {\footnote {Basically $w_0$ it is unconstrained in the phantom direction (with CMB only) upto the
prior range considered.} whereas the non-CMB data
brings it closer to the Cosmological Constant ($w_{0} = -1$). For the combined datasets, 
we find that the mean $w_0$ comes close to the
Cosmological Constant ($w=-1$) but still stays in the phantom region. 

\begin{figure*}[!htb]
 
\begin{center} 
\resizebox{200pt}{180pt}{\includegraphics{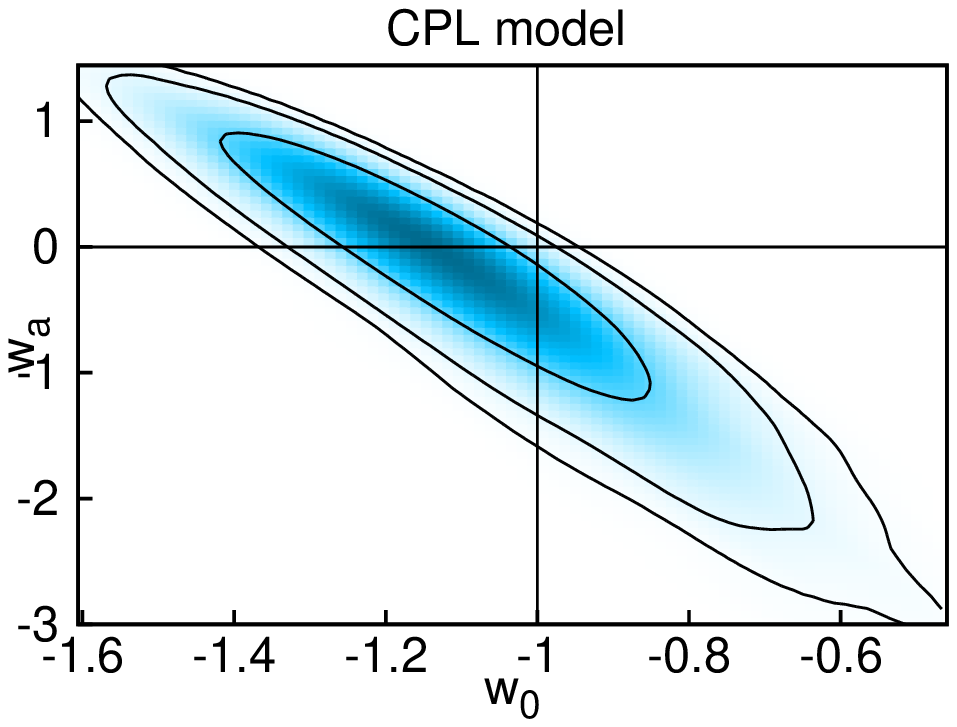}}
\resizebox{200pt}{180pt}{\includegraphics{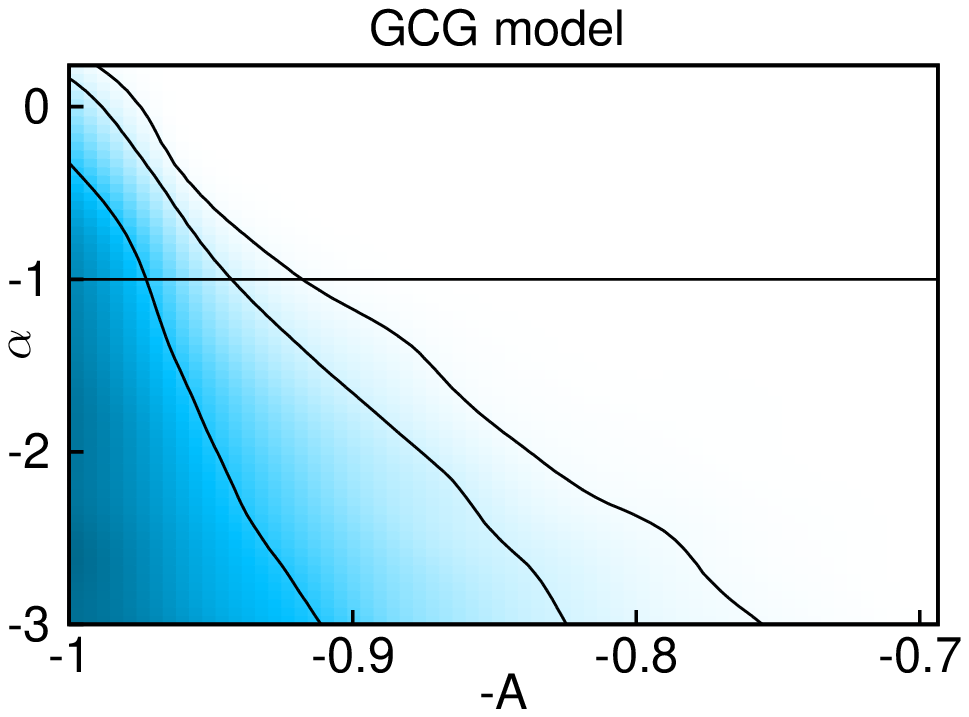}} 
\end{center}
\caption{~\label{fig:2dcntrswwa}Contour plots in the $w_{0}-w_{\rm a}$ plane for CPL and $A-\alpha$ plane for the GCG parametrization.}
 
 \end{figure*}

One can argue that there remains a
tension between CMB and non-CMB data which questions
the effectiveness of a joint analysis of CMB plus non-CMB data together in
future. However, it is hard to pull out a decisive argument from these plots
whether and to what extent the Cosmological Constant is consistent with the data. For a better understanding, we need to look at the 2D marginalized contours of dark energy equation of state parameters.

In Figure~\ref{fig:2dcntrswwa}, we show the marginalized 2D contour plots in
$w_{0}-w_{\rm a}$ and $A-\alpha$ parameter plane for CPL and
GCG parametrization respectively. The CPL case confirms the results earlier
obtained by the Planck collaborations ~\footnote{A tiny deviation from the
Planck result is mainly due to use of all the non-CMB data together in our
analysis}.
It shows that the Cosmological Constant behavior ($w_{0} = -1, w_{\rm a} = 0$) is
disallowed at $1\sigma$ confidence level. Moreover the region $w_{0}> -1$ and
$w_{\rm a} > 0$ is highly constrained even at $3\sigma$ confidence level showing that it is very unlikely that dark energy behaved in non-phantom manner at all epochs.

For GCG parametrization (which is valid only for non-phantom region), as seen in  Figure~\ref{fig:2dcntrswwa}, from the 
relative shaded region below and above the $\alpha=-1$ horizontal line we find that the
 thawing behavior ($\alpha <-1$) is more probable than the freezing behavior
($\alpha > -1$) for dark energy. We would like to highlight that CMB and non-CMB data, in this
context of GCG models, qualitatively distinguishes the equation of state of dark energy. 
For example, while the CMB data constrains the value of $\alpha$ from above, non-CMB can not provide a bound there. On the other hand, in
constraining $-A$, the opposite happens. Since, non-CMB data (mainly SN) provides the most stringent constraints at the present epoch it can constrain $A$ much better than CMB data from Planck.
However SN data does not provide any information beyond $z \sim 1.4$ and therefore can not constrain $\alpha$ as good as CMB data. Note that for $1+\alpha >0$ the denominator 
of Eq.~\ref{eq:gcgeq}) diverges at high redshift (small $a$) and makes $w\simeq0$ (regardless of $A$) which is not supported by CMB data. This result clearly shows the sensitivity of two different probes toward 
two different parameters in the dark energy equation of state. Here, unlike the CPL and SS parametrizations, we can argue that CMB and non-CMB data may go through a joint analysis in 
order to obtain a tighter constraint on the dark energy equation of state. 
For both CPL and SS parametrization (that allow phantom behavior), phantom type equation of state is preferred for
CMB+non-CMB data. Hence, irrespective of the equation of state
parametrization with different number of parameters, phantom is preferred
behavior for combined data. We should point out that allowing the phantom region in the parameter space brings in tension between CMB and non-CMB (specifically SN data); note that 
there is no tension between them when we just concentrate on the non-phantom scenarios. 
At this point we would like to stress on the issue that if the two data sets do not have any systematic 
effects and we claim that a reasonable theory 
should explain the cosmology obtained from different observational data to similar extent,
our results highlights the interesting possibility of exploring and expanding theoretical ideas 
of dark energy which can fundamentally supports an equation of state 
which evolves from a phantom nature at high redshifts towards a Cosmological Constant 
behavior at low redshifts may resolve the tension.

%

\begin{figure*}[!t]

\begin{center} 
\resizebox{166pt}{140pt}{\includegraphics{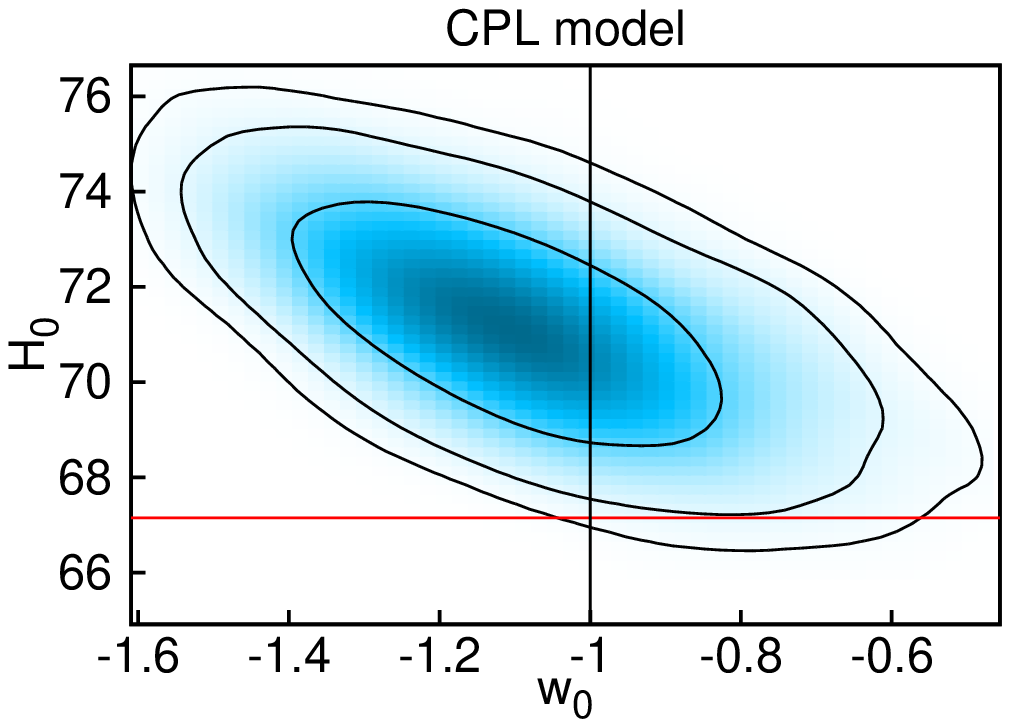}} 
\hskip -35 pt\resizebox{166pt}{140pt}{\includegraphics{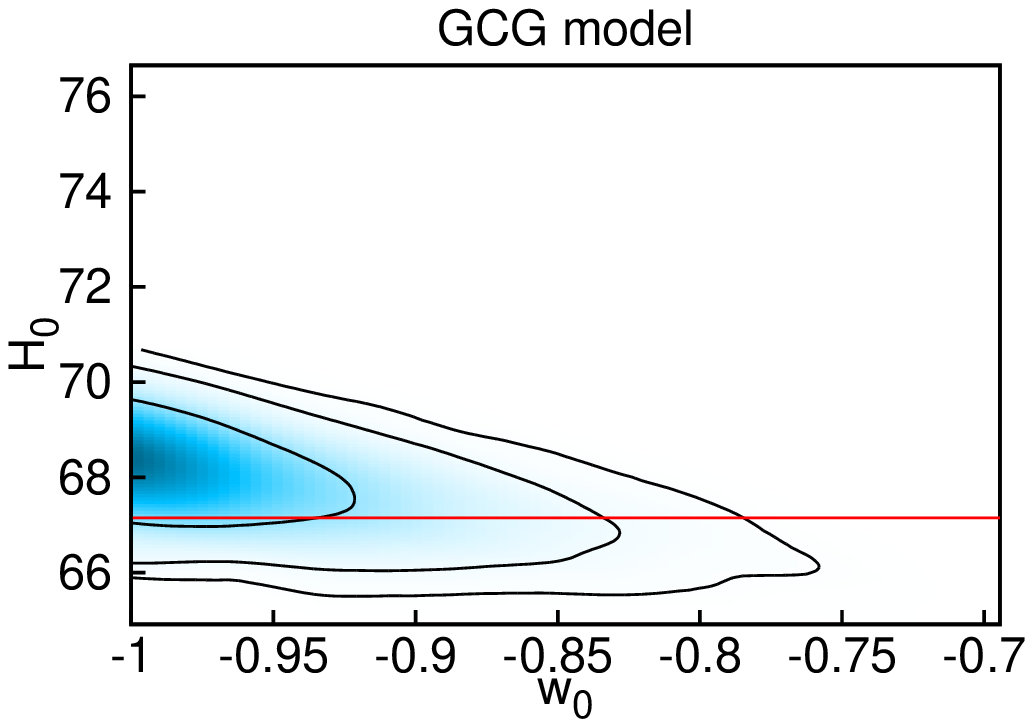}} 
\hskip -35 pt\resizebox{166pt}{140pt}{\includegraphics{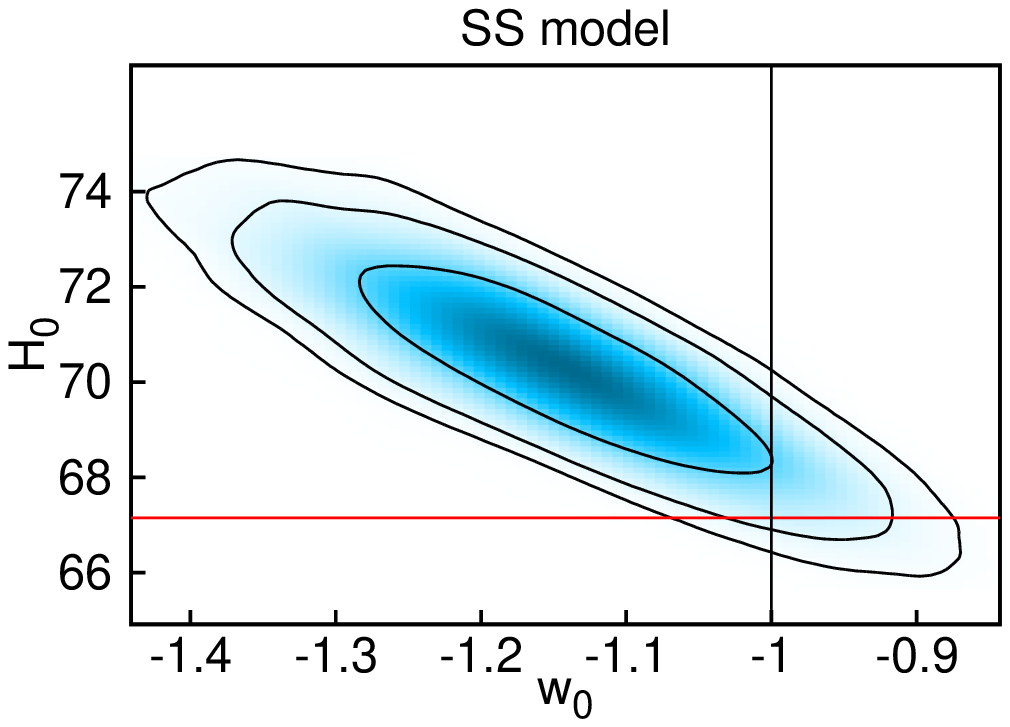}} 
\end{center}
\caption{~\label{fig:2dcntrsh0w}Contour plots in $w_{0}-H_{0}$ parameter plane
for CPL (left), GCG (middle) and SS (right) parametrization. The
red line represents the best-fit value for $H_{0}$ obtained from Planck for
$\Lambda$CDM case.}
 \end{figure*} 

\begin{figure*}[!htb]

\begin{center} 
\resizebox{168pt}{140pt}{\includegraphics{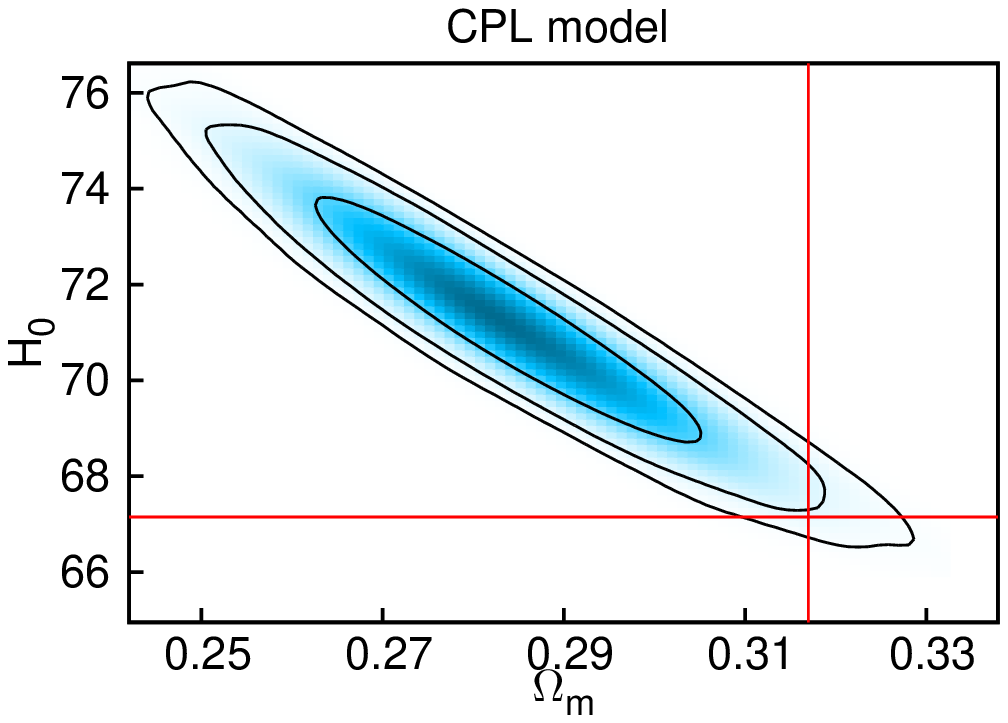}} 
\hskip -38 pt\resizebox{168pt}{140pt}{\includegraphics{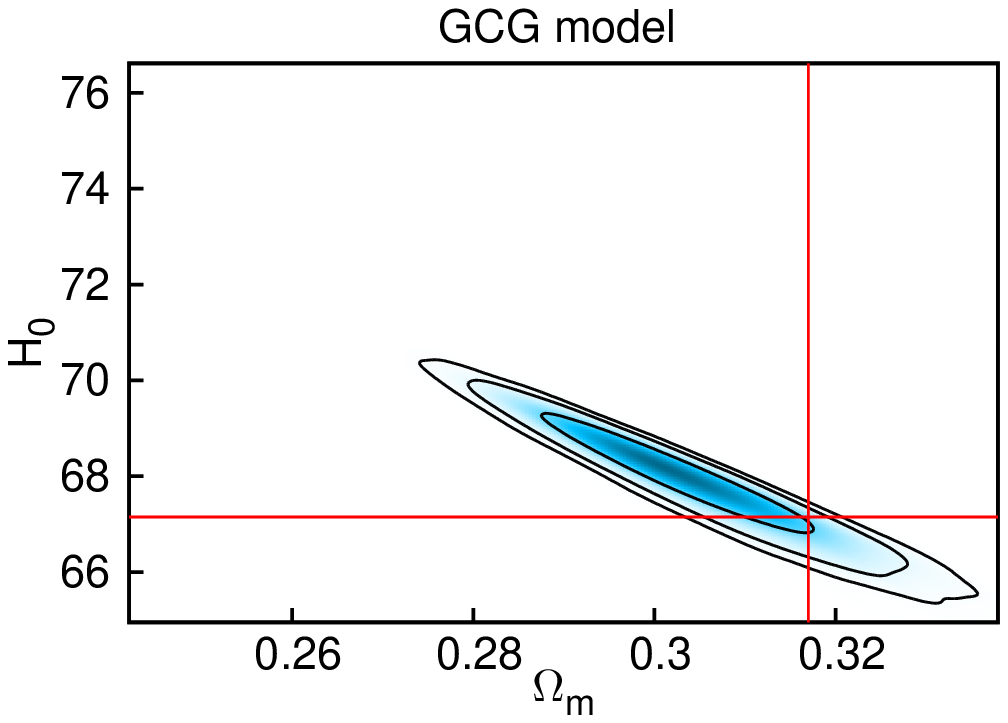}} 
\hskip -38 pt\resizebox{168pt}{140pt}{\includegraphics{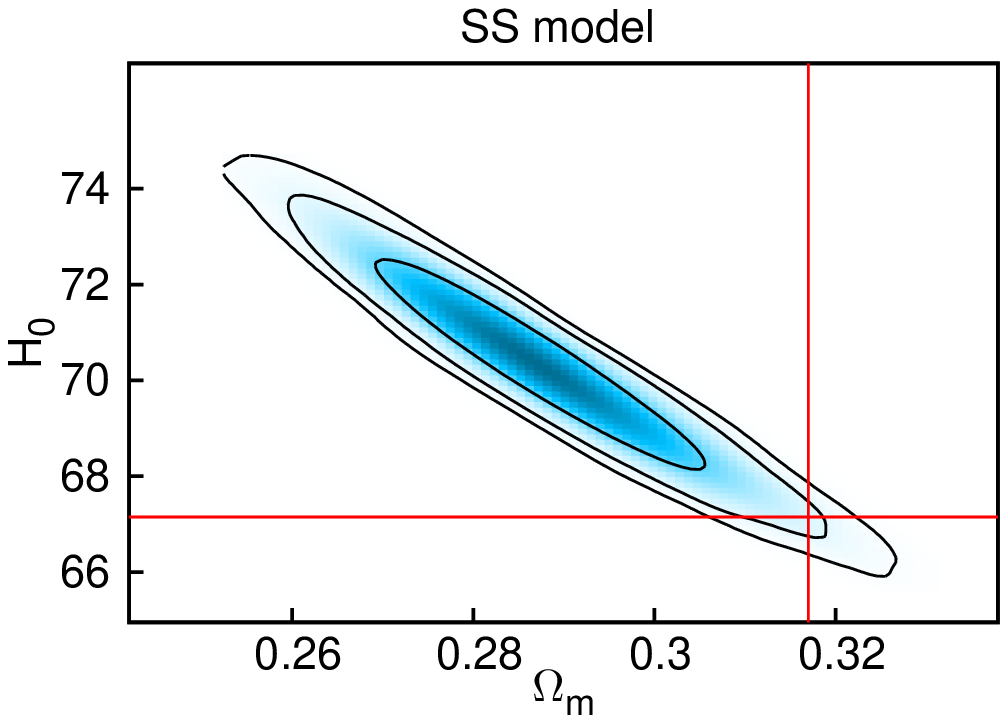}} 

\end{center}
\caption{~\label{fig:2dcntrsommh0}Contour plots in $\Omega_{\rm m}-H_{0}$
parameter plane for CPL (left), GCG (middle) and SS (right) parametrization.
The red lines represents the best-fit values for $H_{0}$ and $\Omega_{\rm m}$
obtained from Planck for $\Lambda$CDM model.}
 \end{figure*}

Next, we look at 2D confidence regions for other cosmological parameters.
In Figure~\ref{fig:2dcntrsh0w}, we show the confidence contours in the
$w_{0}-H_{0}$ plane for all three dark energy parametrizations. 
The Planck best-fit measurements for $H_{0}$ for a concordance $\Lambda$CDM
model is shown by the horizontal red line. The vertical black line represents Cosmological Constant.
In Figure~\ref{fig:2dcntrsommh0}, we show the confidence contours in the
$\Omega_{\rm m}-H_{0}$ parameter plane for all three
parametrization; the red lines show the best-fit values for $\Omega_{\rm m}$ and
$H_{0}$ as measured by Planck for a concordance $\Lambda$CDM model. At this stage, let us mention 
that overplotting the Planck best-fit values does not imply any consistency or inconsistency of
different datasets. Our aim is to address the issue of phantom behavior of the dark energy, i.e whether it is allowed and, if allowed, to what extend does that affect the constraints on other cosmological parameters?


A look at Figures~\ref{fig:2dcntrsh0w} and ~\ref{fig:2dcntrsommh0} show that if we allow phantom equation of state ({\it i.e.} for CPL and SS parametrizations), the best fit cosmology shifts to a higher value of $H_0$ and lower value
of $\Omega_{\rm m}$. This shift leaves the base
$\Lambda$CDM values measured by Planck outside 2$\sigma$ confidence limit for dark energy models captured by the CPL parametrization
and at the border of 2$\sigma$ contour in case of SS parametrized models. However, for dark energy models described by the GCG parametrization - which does not allow phantom models - the Planck $\Lambda$CDM values are 
at the border of the 1$\sigma$ C.L.. To summarize, Planck measurements of
high $\Omega_{\rm m}$ and low $H_{0}$ values for $\Lambda$CDM model are
consistent with measurements of these two parameters using
both CMB + non-CMB data {\it if} we restrict ourselves only to non-phantom models like
GCG. However, tension arises when we allow phantom behavior,
as the CMB and non-CMB data drag the equation of state in two different
directions (i.e., both phantom and non-phantom) and the phantom region provides better fit to the joint likelihood of CMB and non-CMB by 
dominantly better fitting Planck data (in the joint analysis).~\footnote{Note, again,
 that due to the phantom behavior, a higher value of $H_0$ is favored
involving Planck data which is in better agreement with HST.}

\begin{figure*}[!htb]
\begin{center} 
\hskip -20 pt\resizebox{150pt}{140pt}{\includegraphics{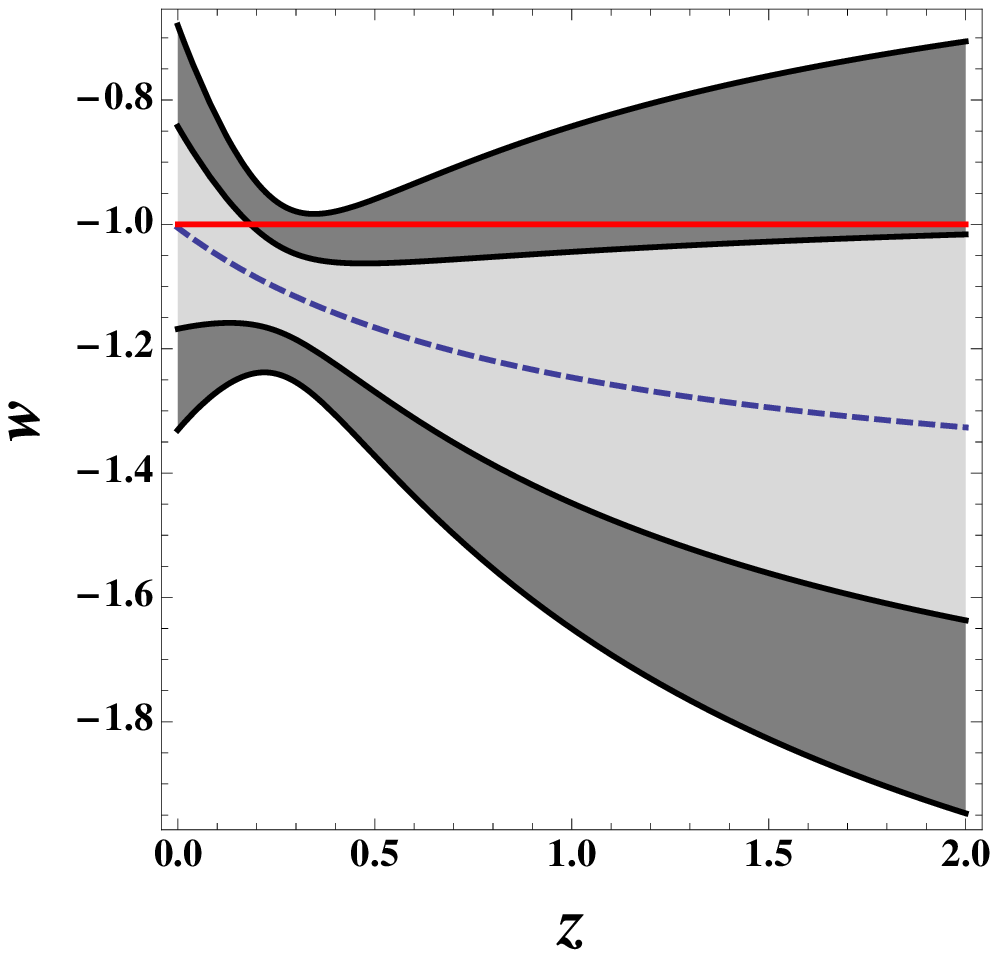}} 
\resizebox{150pt}{140pt}{\includegraphics{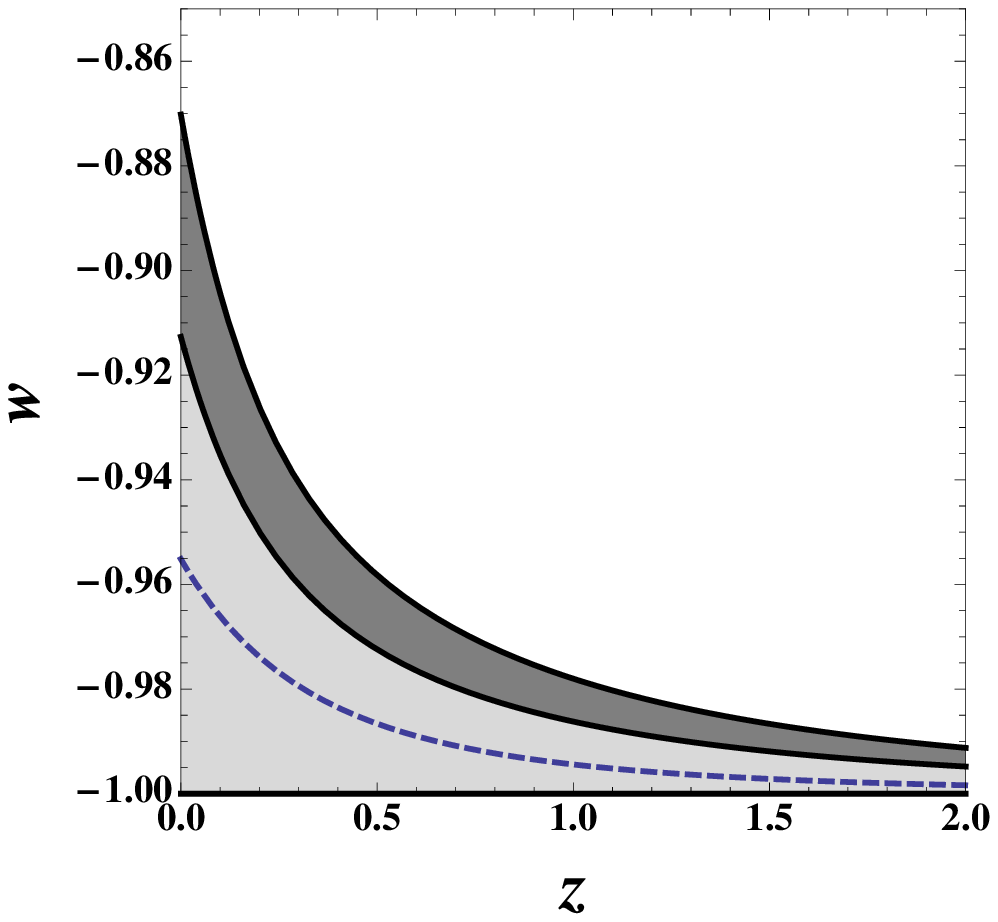}} 
\resizebox{150pt}{140pt}{\includegraphics{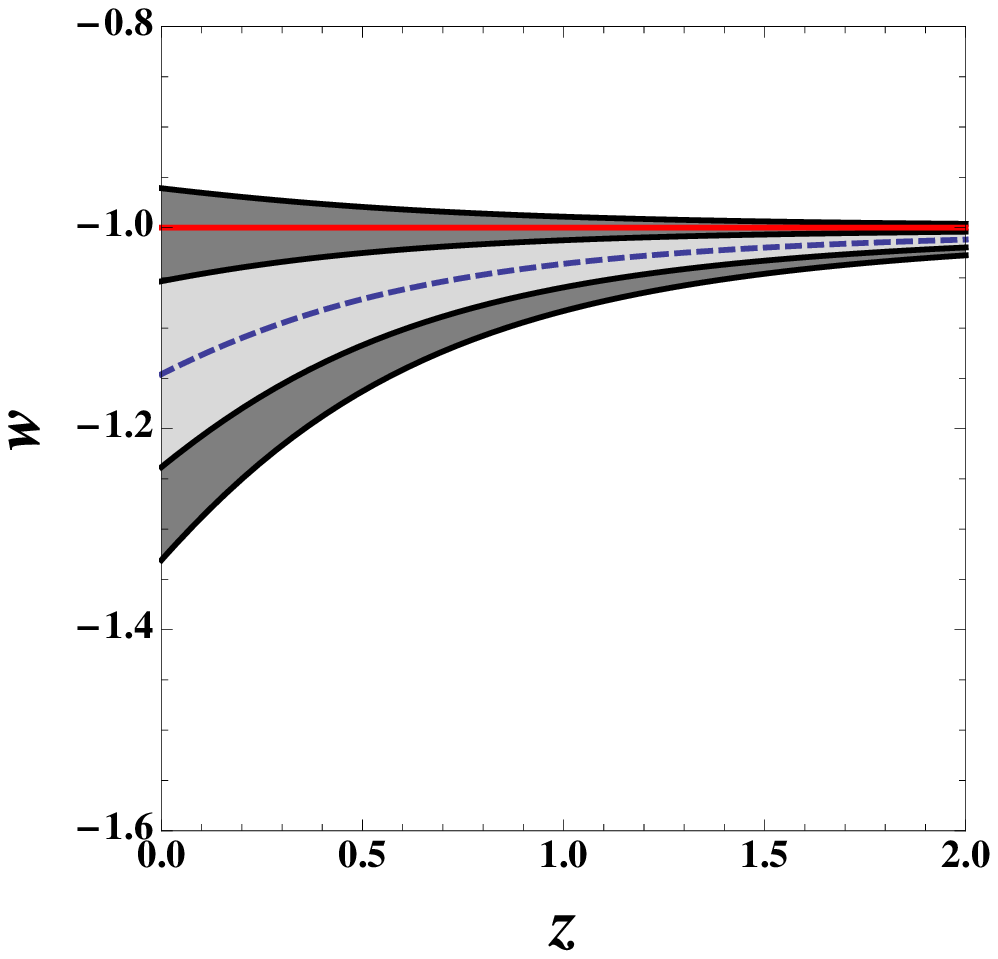}} 

\end{center}
\caption{~\label{fig:evolution}Behavior of equation of state $w$ as a function
of redshift $z$ for CPL (left), GCG (middle) and SS
(right) parametrization for $1-\sigma$ and $2-\sigma$ confidence level. The red and blue lines correspond to $w=-1$ and the mean $w$ respectively. }
 \end{figure*}

The behaviors of the equation of state as
a function of redshift at $1\sigma$ and $2\sigma$ confidence levels,
for the three parametrizations, are shown
in Figure~\ref{fig:evolution}. \\

\noindent A) It is apparent that SS parametrization (right panel in the figure)
constrains the equation of state to evolve very closely to the $w=-1$; the 1$\sigma$ region for $w$ is always less than but within $\sim 5\%$ to the Cosmological Constant at all epochs . This is
expected as this parametrization represents thawing class scalar field models
with small deviations from Cosmological Constant. Thus,
the non-phantom behavior is  not allowed at $1\sigma$ confidence level. 
However, at $2\sigma$, the dark energy behavior is consistent with a 
Cosmological Constant. Note, that the mean $w$ is always phantom and can provide reasonable
 deviation from Cosmological Constant behavior at the present epoch. \\
 
\noindent B) The CPL parametrization (left panel in the figure) has similar behavior and dark energy appears to be phantom at more than  
1$\sigma$ level beyond $z > 0.2$. Only at $z<0.2$, we get $w=-1$ line within the 1$\sigma$ bound. For $z < 0.2$, the mean equation of state 
touch $w=-1$ from below. In this parametrization, mean dark energy equation of state starts as a Cosmological Constant at only $z=0$ and 
quickly deviates from it to become phantom like as redshift increases. $w$ is best constrained at roughly around $z\sim0.3$.
Note that dark energy with $w\ge-1$ lies outside the inner grey region ($1\sigma$) in the range ($0.2 < z < 2$). Moreover, 
it needs the equation of state parameters to be extremely fine tuned to have $w\ge-1$ around $z\simeq0.3$ and still be inside the $2\sigma$ band.
Contrary to the SS parametrization, the CPL parametrization poorly 
constrain $w$ at higher z, whereas at the present epoch both constrain $w$ at roughly $20-30\%$ level.
This behavior of the equation of state is consistent with what we observe in the
$w_{0}-w_{\rm a}$ confidence plane as discussed earlier. 
To summarize, $w$ is consistent with $\Lambda$CDM only at $z=0$ but is increasingly phantom like with increasing  redshift \footnote{Similar conclusions 
about the phantom behavior of dark energy has recently reported by
the  Pan-STARRS1 survey ~\cite{Rest:2013bya}.}. 

 It is easy to understand the different behavior $w$ for SS and CPL prametrizations. 
SS parametrization is designed to follow $w=-1$ in early epoch and has flexibilities only at low redshifts. CPL, on the other hand, 
can allow more scenarios with one more degree of freedom. The important similarity 
between them is both the solution permit larger area in phantom region. 
Note that the SS mean $w$ is allowed within the CPL 1$\sigma$ bounds. The tighter Constraints on $w$ for SS model at high redshift 
is also reflects the inability of SS model to depart from $w=-1$ at early epochs.\\

\noindent C) The mean equation of state for the GCG
parametrization (middle panel in the figure) shows that $w$ reaches the value $\approx -0.96$ at present epoch, 
having started close to or at $w=-1$ at high redshifts. The error bars 
clearly point towards thawer class of models and hence provides an important constraint for
scalar field models for dark energy. For this class of models, in the entire redshift range, 
the deviation of the mean $w$ from a Cosmological Constant is far less than what are found for the other 
classes of models.\\

To summarize, one of our central results is that if one allows for phantom
behavior in the dark energy equation of state, the phantom region
provides a better fit to the combined CMB and non-CMB data. Provided that CMB
and non-CMB joint analysis does not impose systematic errors as has been
discussed before, {\it our results can therefore be thought of as an invitation to
construct models of
dark energy which lead to phantom behavior}, at least at the scales probed
by the Planck and non-CMB observations.
%

\section{Conclusions}
The post-Planck era has seen cosmological parameters best constrained till date using a CMB and a host of non-CMB measurements. This raises the possibility of precision determination
 of the nature of dark energy. 
In this paper we do a detail
investigation with such an aim in mind \footnote{After we submitted our paper to the arXiv, a similar study and very similar
conclusions were shown in ~\cite{Novosyadlyj:2013nya}}. Currently, almost all the cosmological constraints on
dark energy is based on a single parametrization, e.g.
the CPL parametrization. However, theoretical models of dark energy abound in the literature, which leads to possibilities of having models that may not be fairly
represented by such a parametrization. Question remain as to
whether there are possible dark energy evolutions that one misses
using the CPL parametrization.

To address the above mentioned question we work with two other parametrizations, namely SS and GCG, apart from CPL. The SS model, which
describes the dynamical equation of state with a single parameter, allows
both phantom and non-phantom behavior and probes
deviations close to the Cosmological Constant. The GCG model, on the other hand, represents only non-phantom models
allowing {\it only} positive kinetic energies of underlying scalar field
model and provides a clear distinction between tracker and thawer models. Finally, while the CPL parametrization was proposed as a
phenomenological form for the equation of state of dark energy, both the SS and
GCG parametrizations were obtained from a specific field theory Lagrangian under certain conditions \cite{gcg,ss2}. Hence, the three parametrizations together probe maximally possible parameter space for dark energy models.

Having the three parametrizations in hand, we use CMB and non-CMB data in a 
separate and combined analysis to look at dark energy behaviors. The main results of this study are summarized below:
\begin{itemize}
\item 
We find that {\it if} we allow phantom behavior of dark energy, irrespective of misgivings to its use due to the negative kinetic energy of scalar fields,
then CMB data favors phantom over non-phantom behavior with high confidence. 
On the other hand non-CMB data prefer non-phantom behavior for every parametrization considered. 
Once phantom behavior is allowed ({\it i.e.} in CPL \& SS parametrizations), the combined CMB + non-CMB data allows regions such that the Cosmological Constant ($w=-1$) is pushed outside than 1$\sigma$
confidence level contour. Any tension between CMB and non-CMB datasets may be attributed to unknown systematics or the lack of a better theory/parametrization of the dark energy equation of state.

\item Using the correlation between the equation of state parameters, we
reconstruct the late time ({\it i.e.} $z<2$) evolution of dark energy. For the SS
models, the $w=-1$ line stays outside 1$\sigma$ reconstructed band of the evolution
history of dark energy for the entire redshift range. In the case of CPL model, 
the $w=-1$ line stays outside 1$\sigma$ region for $z > 0.2$.

\item The GCG parametrization, which are theoretically constructed so as not 
to allow phantom models, shows consistency between CMB and non-CMB data although 
they have worse likelihood than other parametrizations. 
It is found that the CMB and non-CMB observations are separately sensitive
to the two parameters of the GCG parametrization and that the joint
constraint is consistent with the Cosmological Constant.
For these models, the cosmological parameters too are consistent with base Planck
best-fit measurements. 
Allowing phantom behavior of dark energy reveals tension between CMB and non-CMB (mainly SNe) 
data since CMB data (from Planck) prefers the phantom dark energy behavior and on the other hand non-CMB 
data prefers cosmological constant and non phantom behavior.

\item From the results obtained with the three parametrizations, it comes out that for scalar field dark energy models, a thawing behavior is more
probable than the freezing behavior. This is most clearly been demonstrated in GCG
models. Such an outcome is particularly interesting in the
context of recent construction of axionic quintessence model in string theory
\cite{pst} which is of thawing nature \cite{asthaw}.

\item
The constraints on dark energy, coming out of a joint analysis of all available data, differ from model to model. Not only does the mean $w$ depends on the parametrization of choice but also the error bar on the mean has different behavior. For SS and GCG parametrization, the nature of dark energy is best constrained at high redshifts; however, for the CPL parametrization the best constraints come in the redshift range of $\approx 0.2--0.3$.

\item It has been already noticed in the cosmology community that the Planck measurement for
the parameters $\Omega_{\rm m}$ and $H_{0}$ for the
$\Lambda$CDM model is in tension with the similar measurements by the HST. We
find that for
SS and CPL parametrizations, where we allow phantom, a better fit to the data comes with a large value of $H_0$ which helps to agree better with the
HST data. In fact, the improvement in the total $\chi^2_{\rm eff}$ for the CPL and SS model over $\Lambda$CDM and GCG is driven by the ability to have a higher $H_0$ albeit along with the associated preference for phantom behavior. 
However, in these parametrizations, the phantom effect drags
the background cosmological parameter space in such
a way that the corresponding best-fit base model and that from Planck becomes 2$\sigma$ away from each other.
This implies that if we include phantom model
in our theoretical framework in any cosmological calculations, then we are not
allowed to work with Planck values for the base  $\Lambda$CDM background
parameters. However, GCG model, the only pure non-phantom model, shows that,
within the allowed range of dark energy equation of state,
the background cosmological parameters are in
agreement with Planck values for $\Lambda$CDM.

Also, note that when we allow phantom behavior, the $H_0$ becomes highly degenerate with dark energy equation of state in case of CMB only measurements. Other parametrizations which allow non phantom behaviors only does not exhibit similar behavior; the $H_0$ errors for CPL is $\approx 5$ times larger compared to GCG for CMB only, even though both have the same number of degrees of freedom.

\end{itemize}

The conclusions drawn above, in the current work, comes from a joint analysis of CMB and non-CMB data using different evolving dark energy models having different 
parametrization of the equation of state of dark energy. In a detailed analysis of SNe data (along with other datasets),the PAN-STARRS1 survey
 ~\cite{Rest:2013bya} recently found hints for similar phantom behavior of dark energy, although using a constant equation of state.
 Their value of w is inconsistent with the Cosmological Constant value of -1 at  $> 2\sigma$ level (if used along with Planck data) or at $< 2\sigma$ (if used along with WMAP9 data).
 Recently, in an arXiv submission ~\cite{Novosyadlyj:2013nya} following our work, it was also reported that for dataset Planck+HST+BAO+SNLS3 
 the $\Lambda$CDM model is just outside the 2$\sigma$ confidence regime, while for the dataset 
 WMAP9+HST+BAO+SNLS3 the $\Lambda$CDM model is 1$\sigma$ away from the best fit. All these works nicely complements each other.
 
A very important point, not adequately appreciated in the literature 
(and missed in the two papers ~\cite{Rest:2013bya, Novosyadlyj:2013nya} cited above), 
comes out in Figure~\ref{fig:evolution}. This is the fact that {\it our 
constraints on $w$ and hence the nature of dark energy that we infer from cosmological 
observation depends crucially on the choice of the underlying parametrization of the equation of state.} 
In fact, (any) deviation of $w$ from a Cosmological Constant, with redshift depends on the parametrization - 
whereas for the current available data, both SS and GCG are infers dark energy being close to the Cosmological
Constant at high redshift but `deviating from it in two different directions' when we approach current epoch; on
the other hand CPL is close to Cosmological Constant at current epoch and deviates away at high z. Given this trichotomy,
it is important to do non-parametric reconstruction of $w$ for the total dataset to infer the correct nature of dark energy 
without any prior on form of $w$. 

To summarize once again, we have done the most comprehensive study of dark energy equation of state constraints from a joint analysis of Planck CMB data along with SNe, BAO and $H_0$ data.
A central results of our analysis is that if one allows for phantom
behavior in the dark energy equation of state, the phantom region
provides a better fit to the combined CMB and non-CMB data. 
The tension with $\Lambda$CDM can be due to the even more mysterious nature of dark energy or a combination of chance and systematic errors.
Our results motivate the construction of models of
dark energy which lead to phantom behavior. This means going beyond
standard possibilities for dark energy involving a scalar field with a positive
kinetic energy term only which do not lead to a violation
of the weak energy condition. However, it is well known
that in consistent theories of gravity, like string theory,
the weak energy condition and also the null energy condition can be violated due to the presence of  higher derivative
corrections ~\cite{Creminelli:2008wc}. We leave it as an intriguing question to the reader as to whether such violations can be
used to construct a model of dark energy which would fit the data
better than say a positive  Cosmological Constant.

\section*{Acknowledgments}
The authors would like to thank Sandip Trivedi for motivation, discussions and support throughout the project.
D.K.H wish to acknowledge support from the Korea Ministry of Education, Science
and Technology, Gyeongsangbuk-Do and Pohang City for Independent Junior Research Groups at
the Asia Pacific Center for Theoretical Physics. We also acknowledge the use of
publicly
available CAMB and CosmoMC in our analysis. A.A.S. acknowledges the funding from SERC, Dept. of Science and Technology, Govt. of India through the research
project SR/S2/HEP-43/2009. SP thanks ISI, Kolkata for support through research
grant. This project is part of the {\sf Dark Universe Initiative} program.



\begin{thebibliography}{99}

\bibitem{review}
E. J. Copeland, M. Sami and S. Tsujikawa, Int.~J.~Mod.~Phys. D {\bf 15}, 1753
(2006);
Miao Li, Xiao-Dong Li, Shuang Wang,  arXiv:1103.5870;
V.~Sahni and A.~A.~Starobinsky, Int.\ J.\ Mod.\ Phys.\  D {\bf 9}, 373 (2000);
S.~M.~Carroll, Living Rev.\ Rel.\  {\bf 4}, 1 (2001);
P.~J.~E.~Peebles and B.~Ratra, Rev.\ Mod.\ Phys.\  {\bf 75}, 559 (2003);
T.~Padmanabhan, Phys.\ Rept.\  {\bf 380}, 235 (2003).


\bibitem{sn}
A.~G.~Riess, et al.,  Astron.~J. {\bf 116}, 1103, (1998);
A.~G.~Riess, et al., Astrophys.~J.,, {\bf 607}, 665, (2004);
S.~Perlmutter, et al.,  Astrophys.~J., {\bf  517}, 565, (1999); 
J~L.~Tonry, et al., Astrophys.~J., {\bf 594}, 1,   (2003);
R.~A.~Knop, et al.,  Astrophys.~J., {\bf  598}, 102, (2003);
R.~Amanullah, et al., Astrophys.~J.,  {\bf 716}, 712, (2010);
N.~Suzuki et~al., Astrophysics J.~ {\bf 746}, 85 (2012).


\bibitem{komatsu} E.~Komatsu et al., Astrophys. J. Suppl. {\bf 192}, 18 (2011).


\bibitem{Planck}
See, {\tt
http://www.sciops.esa.int/index.php?project=planck\&page=Planck\_Collaboration}

\bibitem{Eisenstein} D.~Eisenstein,  Astophys.~J., {\bf 633}, 560 (2005).

\bibitem{Kachru:2003sx} 
  S.~Kachru, R.~Kallosh, A.~D.~Linde, J.~M.~Maldacena, L.~P.~McAllister and S.~P.~Trivedi,
  JCAP {\bf 0310}, 013 (2003)
  [hep-th/0308055]



\bibitem{quint}
B. Ratra and P.J.E. Peebles,
Phys.~Rev.~D,  {\bf 37}, 3406 (1988);
M.S. Turner and M. White,
Phys.~Rev.~D, {\bf 56}, R4439 (1997)
R.R. Caldwell, R. Dave, and P.J. Steinhardt,
Phys.~Rev.~Lett.  {\bf 80}, 1582 (1998);
A.R. Liddle and R.J. Scherrer,
Phys.~Rev.~D, {\bf 59}, 023509 (1999);
P.J. Steinhardt, L. Wang, and I. Zlatev,
Phys.~Rev.~D, {\bf 59}, 123504 (1999);
R.~J.~Scherrer and A.~A.~Sen, Phys.~Rev.~D, {\bf 77}, 083515 (2008).

\bibitem{kess}
C. Armendariz-Picon, T. Damour, and V. Mukhanov, Phys.~Lett.~B {\bf 458}, 209
(1999);
J. Garriga and V.F. Mukhanov,
Phys.~Lett.~B {\bf 458}, 219 (1999);
T. Chiba, T. Okabe, M. Yamaguchi,
Phys.~Rev.~D, {\bf 62}, 023511 (2000);
C. Armendariz-Picon, V. Mukhanov, and P.J. Steinhardt,
Phys.~Rev.~Lett. {\bf 85}, 4438 (2000);
C. Armendariz-Picon, V. Mukhanov, and P.J. Steinhardt,
Phys.~Rev.~D, {\bf 63}, 103510 (2001);
T. Chiba, Phys.~Rev.~D, {\bf 66}, 063514 (2002);
L.P. Chimento and A. Feinstein, Mod.~Phys.~Lett.~A {\bf 19}, 761 (2004);
L.P. Chimento, Phys.~Rev.~D, {\bf 69}, 123517 (2004);
R.J. Scherrer, Phys.~Rev.~Lett. {\bf 93}, 011301 (2004).

\bibitem{phant}R.~R.~Caldwell, Phys.~Lett.~B {\bf 545}, 23 (2002). 

\bibitem{tach}
J.~S.~Bagla, H.~K.~Jassal and T.~Padmanabhan, Phys.~Rev.~D {\bf 67}, 063504
(2003);
A.~A.~Sen, JCAP, {\bf 0603}, 010, (2006).

\bibitem{gcg}
M.C. Bento, O. Bertolami, and A.A. Sen,Phys.~Rev.~D, {\bf 66}, 043507 (2002)

\bibitem{ss1}
A.~A.~Sen and R.~J.~Scherrer, Phys.~Rev.~D, {\bf 72}, 063511 (2005).

\bibitem{Chevallier:2000qy} 
  M.~Chevallier and D.~Polarski,
  Int.\ J.\ Mod.\ Phys.\ D {\bf 10}, 213 (2001)
  [gr-qc/0009008].



\bibitem{Linder:2002et} 
  E.~V.~Linder,
  Phys.\ Rev.\ Lett.\  {\bf 90}, 091301 (2003)
  [astro-ph/0208512].


\bibitem{Scherrer:2007pu} 
  R.~J.~Scherrer and A.~A.~Sen,
  Phys.\ Rev.\ D {\bf 77}, 083515 (2008)
  [arXiv:0712.3450 [astro-ph]].

\bibitem{ss2}
 R.~J.~Scherrer and A.~A.~Sen
 Phys.\ Rev.\ D {\bf 78}, 067303 (2008)

\bibitem{Ali:2009mr} 
  A.~Ali, M.~Sami and A.~A.~Sen,
distinguish it from quintessence ?,''
  Phys.\ Rev.\ D {\bf 79}, 123501 (2009)
  [arXiv:0904.1070 [astro-ph.CO]].

\bibitem{pandadbi}
E.~Bergshoeff, M.~de.~Roo, T.~de~Wit, E.~Eyras and S.~Panda, JHEP {\bf 0005} 009(2000).
  

\bibitem{Bento:2003dj} 
  M.~C.~Bento, O.~Bertolami and A.~A.~Sen,
  Gen.\ Rel.\ Grav.\  {\bf 35}, 2063 (2003)
  [gr-qc/0305086].

\bibitem{Bento:2002uh} 
  M.~C.~Bento, O.~Bertolami and A.~A.~Sen,
  astro-ph/0210375.

\bibitem{Bento:2002yx} 
  M.~C.~Bento, O.~Bertolami and A.~A.~Sen,
  Phys.\ Rev.\ D {\bf 67}, 063003 (2003)
  [astro-ph/0210468].

\bibitem{Bento:2003we} 
  M.~C.~Bento, O.~Bertolami and A.~A.~Sen,
  Phys.\ Lett.\ B {\bf 575}, 172 (2003)
  [astro-ph/0303538].

\bibitem{Thakur:2012rp}
  S.~Thakur, A.~Nautiyal, A.~A.~Sen and T RSeshadri,
  Mon.\ Not.\ Roy.\ Astron.\ Soc.\  {\bf 427} (2012) 988
  [arXiv:1204.2617 [astro-ph.CO]].

\bibitem{Shafieloo:2007cs} 
  A.~Shafieloo,
  Mon.\ Not.\ Roy.\ Astron.\ Soc.\  {\bf 380}, 1573 (2007)
  [astro-ph/0703034 [ASTRO-PH]].

\bibitem{Shafieloo:2012yh} 
  A.~Shafieloo,
JCAP {\bf 1208}, 002 (2012)
  [arXiv:1204.1109 [astro-ph.CO]].

\bibitem{Shafieloo:2009ti} 
  A.~Shafieloo, V.~Sahni and A.~A.~Starobinsky,
  Phys.\ Rev.\ D {\bf 80}, 101301 (2009)
  [arXiv:0903.5141 [astro-ph.CO]].

\bibitem{recent}
Jun-Qing~Xia, Hong~Li and Xinmin~Zhang, Phys. \ Rev.\ D {\bf 88}, 063501
(2013);Valeria~Pettorino, arXiv:1305.7457 [astro-ph.CO];
Valentina Salvatelli and Andrea Marchini, Phys. \ Rev. \ D {\bf 88}, 023531
(2013); 
  E.~Macaulay, I.~K.~Wehus and H.~K.~Eriksen,
  arXiv:1303.6583 [astro-ph.CO];
 C.~Cheng and Q.~-G.~Huang,
  arXiv:1306.4091 [astro-ph.CO].
\bibitem{pst}
Sudhakar~ Panda, Yoske~Sumimoto and Sandip~P.~Trivedi, Phys.\ Rev.\ D {\bf 83},
083506 (2011).

  
\bibitem{Kamenshchik:2001cp}
  A.~Y.~.Kamenshchik, U.~Moschella and V.~Pasquier,
  Phys.\ Lett.\ B {\bf 511} (2001) 265
  [gr-qc/0103004].


\bibitem{bilic}N.~Bilic, G.~B.~Tupper, and R.~D.~Viollier, Phys. Lett. B {\bf
535}, 17 (2002);

\bibitem{Hinshaw:2012aka} 
  G.~Hinshaw {\it et al.}  [WMAP Collaboration],
  arXiv:1212.5226 [astro-ph.CO].

\bibitem{Planck:2013kta} 
  P.~A.~R.~Ade {\it et al.}  [Planck Collaboration],
  arXiv:1303.5075 [astro-ph.CO].

\bibitem{Ade:2013zuv} 
  P.~A.~R.~Ade {\it et al.}  [Planck Collaboration],
  arXiv:1303.5076 [astro-ph.CO].

\bibitem{Percival:2009xn} 
  W.~J.~Percival {\it et al.}  [SDSS Collaboration],
  Mon.\ Not.\ Roy.\ Astron.\ Soc.\  {\bf 401}, 2148 (2010)
  [arXiv:0907.1660 [astro-ph.CO]].

\bibitem{Blake:2011en} 
C.~Blake, E.~Kazin, F.~Beutler, T.~Davis, D.~Parkinson, S.~Brough, M.~Colless
and C.~Contreras {\it et al.},
  Mon.\ Not.\ Roy.\ Astron.\ Soc.\  {\bf 418}, 1707 (2011)
  [arXiv:1108.2635 [astro-ph.CO]].

\bibitem{Anderson:2012sa} 
L.~Anderson, E.~Aubourg, S.~Bailey, D.~Bizyaev, M.~Blanton, A.~S.~Bolton,
J.~Brinkmann and J.~R.~Brownstein {\it et al.},
  Mon.\ Not.\ Roy.\ Astron.\ Soc.\  {\bf 427}, no. 4, 3435 (2013)
  [arXiv:1203.6594 [astro-ph.CO]].

\bibitem{Beutler:2011hx} 
F.~Beutler, C.~Blake, M.~Colless, D.~H.~Jones, L.~Staveley-Smith, L.~Campbell,
Q.~Parker and W.~Saunders {\it et al.},
  Mon.\ Not.\ Roy.\ Astron.\ Soc.\  {\bf 416}, 3017 (2011)
  [arXiv:1106.3366 [astro-ph.CO]].
  
\bibitem{Suzuki:2011hu} 
N.~Suzuki, D.~Rubin, C.~Lidman, G.~Aldering, R.~Amanullah, K.~Barbary,
L.~F.~Barrientos and J.~Botyanszki {\it et al.},
  Astrophys.\ J.\  {\bf 746}, 85 (2012)
  [arXiv:1105.3470 [astro-ph.CO]].

\bibitem{Riess:2011yx} 
A.~G.~Riess, L.~Macri, S.~Casertano, H.~Lampeitl, H.~C.~Ferguson,
A.~V.~Filippenko, S.~W.~Jha and W.~Li {\it et al.},
  Astrophys.\ J.\  {\bf 730}, 119 (2011)
  [Erratum-ibid.\  {\bf 732}, 129 (2011)]
  [arXiv:1103.2976 [astro-ph.CO]].
  
\bibitem{Lewis:1999bs}
  A.~Lewis, A.~Challinor and A.~Lasenby,
  Astrophys.\ J.\  {\bf 538} (2000) 473
  [astro-ph/9911177].

\bibitem{cambsite}
See, {\tt http://camb.info/.}


\bibitem{Lewis:2002ah}
  A.~Lewis and S.~Bridle,
  Phys.\ Rev.\ D {\bf 66} (2002) 103511
  [astro-ph/0205436].
  
\bibitem{cosmomcsite}
See, {\tt http://cosmologist.info/cosmomc/.}

\bibitem{powell}
M.~J.~D.~Powell, Cambridge NA Report NA2009/06, University of Cambridge,
Cambridge (2009).

\bibitem{asthaw}
G.~Gupta, S.~Panda and A.~A.~Sen, Phys.~Rev.~D, {\bf 85}, 023501 (2012)
S.~Kumar, S.~Panda and A.~A.~Sen, Class.~Quant.~Grav. {\bf 30}, 155011 (2013).

\bibitem{Rest:2013bya} 
A.~Rest, D.~Scolnic, R.~J.~Foley, M.~E.~Huber, R.~Chornock, G.~Narayan,
J.~L.~Tonry and E.~Berger {\it et al.},
  arXiv:1310.3828 [astro-ph.CO].


\bibitem{Novosyadlyj:2013nya} 
  B.~Novosyadlyj, O.~Sergijenko, R.~Durrer and V.~Pelykh,
  arXiv:1312.6579 [astro-ph.CO].

\bibitem{Creminelli:2008wc}
  P.~Creminelli, G.~D'Amico, J.~Norena and F.~Vernizzi,
  JCAP {\bf 0902} (2009) 018
  [arXiv:0811.0827 [astro-ph]].


\end{thebibliography}
 \end{document}